\documentclass[%
 amsmath,amssymb,
 aps,
 prl,
 twocolumn,
]{revtex4-2}
\usepackage[english]{babel}
\usepackage{bibunits}
\usepackage[utf8]{inputenc}
\usepackage{amsthm}
\usepackage{mathtools}
\usepackage{physics}
\usepackage{xcolor}
\usepackage{graphicx}
\graphicspath{{./}{sections/Fig1/}{sections/Fig2/}{sections/Fig3/}{sections/Fig4/}{sections/Supplement/}}
\usepackage{tikz}
\usepackage{adjustbox}
\usepackage{multirow}
\usepackage{placeins}
\usepackage[T1]{fontenc}
\usepackage{csquotes}
\usepackage[normalem]{ulem}
\usepackage{enumitem}
\usepackage[pdftex, pdftitle={Bias-preserving cat-cat CNOT gate via vacuum-conditional beam-splitter}, pdfauthor={Yufeng (Bright) Ye, Connor T. Hann, Kyungjoo Noh, Harald Putterman, Ron Belyansky, Arne L. Grimsmo, Oskar Painter}]{hyperref}
\begin{document}
\makeatletter
\newcommand{\combinedalias}[1]{%
  \expandafter\ifx\csname r@#1\endcsname\relax\else
    \expandafter\global\expandafter\let\csname r@sm-#1\expandafter\endcsname\csname r@#1\endcsname
    \expandafter\global\expandafter\let\csname r@main-#1\expandafter\endcsname\csname r@#1\endcsname
  \fi
}
\AfterEndPreamble{%
  \combinedalias{fig:performance}\combinedalias{fig:chiecho}\combinedalias{fig:buildup}\combinedalias{eq:hamiltonian}%
  \combinedalias{fig:qec_gate_input}\combinedalias{fig:control_bin_T1}%
  \combinedalias{sec:T1_comparison}\combinedalias{sec:swap-glossary}\combinedalias{sec:thermal_sensitivity}%
  \combinedalias{sec:sm-toffoli}\combinedalias{sec:sm-drag}\combinedalias{sec:sm-simcnot}\combinedalias{sec:sm-chiecho}\combinedalias{sec:sm-dispecho}\combinedalias{sec:sm-beamsplitter}\combinedalias{sec:sm-beamsplitter-detuned}\combinedalias{sec:sm-simulations}\combinedalias{sec:sm-hardware-realizations}\combinedalias{sec:sm-error-defs}\combinedalias{sec:sm-ctrl-pf}\combinedalias{sec:sm-ctrl-pf-thermal}\combinedalias{sec:sm-kerr-mismatch}\combinedalias{sec:sm-gate-time}\combinedalias{sec:sm-su2}\combinedalias{sec:sm-biasswap}\combinedalias{sec:sm-longitudinal}\combinedalias{sec:sm-idle-free-echo}\combinedalias{sec:sm-displaced-dephasing}\combinedalias{sec:sm-zeno-qec-comparison}\combinedalias{sec:sm-thermal-target-on}\combinedalias{sec:sm-cycling-condition}\combinedalias{sec:sm-hardware-signflip}%
}
\makeatother
\preprint{APS/123-QED}
\title{Bias-preserving cat-cat CNOT gate via vacuum-conditional beam-splitter}

\author{Yufeng (Bright) Ye$^{1}$}
\email[Correspondence email address: ]{brightye@amazon.com}%

\author{Connor T. Hann$^{1}$}
\author{Kyungjoo Noh$^{1}$}
\thanks{Current address: NVIDIA Corporation, USA}
\author{Harald Putterman$^{1}$}
\author{Ron Belyansky$^{1}$}
\author{Arne L. Grimsmo$^{1}$}
\author{Oskar Painter$^{1,2,3}$}

\affiliation{$^1$Amazon Center for Quantum Computing, Pasadena, CA 91125, USA}
\affiliation{$^2$Institute for Quantum Information and Matter, California Institute of Technology, Pasadena, CA 91125, USA}
\affiliation{$^3$Thomas J. Watson, Sr., Laboratory of Applied Physics, California Institute of Technology, Pasadena, CA 91125, USA}

\date{\today}

\begin{abstract}
Cat qubits can exhibit strong noise bias due to their exponentially enhanced bit-flip times and only polynomially reduced phase-flip times with increasing photon number, which makes them attractive candidates for hardware-efficient quantum error correction.
However, it is difficult to maintain this strong noise bias in logical operations such as CNOT gates between cats.
Here, we propose a coherent CNOT gate scheme between two dissipative cats that preserves the exponential noise bias.
The proposed gate relies only on unitary operations, which avoids the non-idealities associated with many existing gate schemes that rely on engineered dissipations.
Assuming good component lifetimes and precise nonlinearity engineering, the proposed gate can enable logical memory in the megaquop regime (logical error rates $< 10^{-6}$) with a distance-$7$ repetition code consisting of 13 cat qubits.
\end{abstract}

\maketitle

\begin{bibunit}[apsrev4-2]

\textit{Introduction}---Quantum error correction (QEC) is essential for useful quantum computation~\cite{Knill1998, Terhal2015}, but standard approaches such as the surface code are estimated to require ${\sim}10^2$ physical qubits per logical qubit~\cite{Gidney2023yoked, Gidney2025rsa}.
Bosonic codes mitigate this overhead by encoding a qubit in the infinite-dimensional Hilbert space of a harmonic oscillator~\cite{GKP2001, Cochrane1999}, with recent experiments demonstrating QEC beyond the break-even point~\cite{Ofek2016, Sivak2023} and concatenation of bosonic qubits~\cite{Putterman2024cqc}.
Among bosonic codes, cat qubits---whether stabilized dissipatively via engineered two-photon loss~\cite{Leghtas2015, Mirrahimi2014} or via Kerr nonlinearity~\cite{Grimm2020}---are especially promising.
Their bit-flip rate is exponentially suppressed with increasing mean photon number $|\alpha|^2$, while the phase-flip rate grows only linearly~\cite{Lescanne2020, Berdou2023, Reglade2024, Putterman2024cat}.
This extreme noise bias enables the use of a simple one-dimensional repetition code to correct the dominant phase-flip errors~\cite{Guillaud2019, Guillaud2021}, in contrast to the two-dimensional surface codes needed for unbiased noise systems.
Resource estimates indicate that ${\sim}10^5$ cat qubits could run algorithms composed of millions of logical gates~\cite{Gouzien2023, Chamberland2022}, with recent LDPC-cat codes promising further overhead reductions~\cite{Ruiz2024}.

However, this overhead advantage holds only if the noise bias is maintained through all gate operations, most importantly the entangling CNOT gate used in syndrome extraction~\cite{Puri2020, Aliferis2008}.
While the syndrome-extraction ancilla can also be a different qubit type, such as a transmon~\cite{Hann2024, Putterman2024cqc}, here we focus on cat ancillas and a bias-preserving cat--cat CNOT gate.
Existing proposals for bias-preserving cat--cat CNOT gates rely on the quantum Zeno effect from the stabilizing two-photon dissipation~\cite{Guillaud2019, Gautier2023zeno}, which imposes an adiabaticity constraint: The gate must be slow compared to the dissipation rate $\kappa_2$, limiting speed and producing residual non-adiabatic phase errors.
Alternative schemes accept partial bias degradation for faster operations~\cite{LeRegent2023}, use time-dependent Hamiltonian engineering for fast bias-preserving gates~\cite{Xu2022}, or combine dissipative and Hamiltonian confinement~\cite{Gautier2022combined}.
All such approaches require engineered dissipation to remain active during the gate.

Our proposal for a bias-preserving CNOT gate between two cat qubits is a unitary (dissipation-free) controlled-SWAP$^2$ (cSWAP$^2$): conditioned on the control state, the target cat is swapped into an auxiliary bin mode and back, acquiring the geometric phase that implements $X$ on the target.
The conditional SWAP is implemented by a vacuum-conditional beam-splitter (VCB) interaction mediated by a cross-Kerr nonlinearity~\cite{Holland2015, Gao2019, Ye2021quarton}.
Without dissipation requirements, this gate is free from non-adiabatic phase-flip errors and can prevent propagation of ancilla bit-flip errors in syndrome extraction.
With realistic parameters, we show that the gate preserves the exponential noise bias for both control and target cats, enabling logical memory with error rates below $10^{-6}$ per cycle in a distance-$7$ repetition code consisting of 13 cat qubits.
Because the gate preserves noise bias on both control and target, it can also be used transversally between repetition-code blocks, which---combined with recent work on algorithmic fault tolerance~\cite{Zhou2024algFT, Cain2024correlated}---can speed up logical computation, and extends naturally to a bias-preserving Toffoli (Sec.~\ref{sm-sec:sm-toffoli} of~\cite{SM}).

\begin{figure}[t]
\includegraphics[width=0.85\columnwidth]{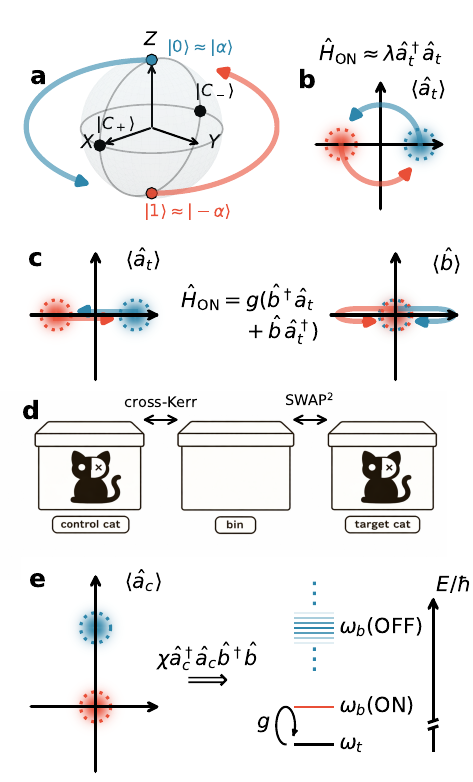}
\caption{\label{fig:gate_concept} Conditional SWAP$^2$ realizes bias-preserving cat-cat CNOT. (a) Bloch sphere of the cat qubit. To remain bias-preserving, an $X$ gate must leave the two-dimensional code space --- its trajectory (colored arcs) exits the Bloch-sphere surface into the larger oscillator Hilbert space, rather than crossing the unprotected equator on the surface. (b) Existing $X$ gates rotate the cat in phase space via a frequency shift Hamiltonian $\lambda\hat{a}^\dagger\hat{a}$. (c) Proposed $X$ gate SWAPs the cat (target mode $\hat{a}_t$) in and out of an auxiliary bin mode ($\hat{b}$) via a beam-splitter Hamiltonian $g(\hat{b}^\dagger\hat{a}_t + \hat{b}\,\hat{a}_t^\dagger)$, completing a SWAP$^2$ geometric phase gate (Sec.~\ref{sm-sec:swap-glossary} of~\cite{SM}). (d) Schematic of the proposed CNOT gate: A cross-Kerr interaction $\chi\hat{a}_c^\dagger\hat{a}_c\hat{b}^\dagger\hat{b}$ between control cat and bin mode conditionally activates the bias-preserving $X$ gate. (e) Energy level diagram of the vacuum-conditional beam-splitter: When control is in $|0\rangle$ (ON), the beam-splitter drive is resonant with the bin-target frequency difference $\omega_b - \omega_t$ and executes a SWAP$^2$; when control is in $|2\alpha_c\rangle$ (OFF), the average cross-Kerr shift $\bar{\Delta} = \chi\langle\hat{n}_c\rangle = 4\chi|\alpha_c|^2$ detunes the bin, pushing the drive off resonance and suppressing the interaction.}
\end{figure}

\textit{Gate concept}---Before presenting our gate, we briefly review why bias-preserving gates are fundamentally difficult for cat qubits.
A bias-preserving CNOT between strictly two-level qubits is conjectured to be impossible~\cite{Aliferis2008, Guillaud2019}. Cat qubits circumvent this by leaving the two-dimensional code space during the gate, as detailed below~\cite{Puri2020, Guillaud2022lectures}.
Cat qubits are stabilized by engineered two-photon dissipation, $\dot{\rho} = \kappa_2\,\mathcal{D}[\hat{a}^2 - \alpha^2]\rho$, which confines the oscillator to the two-dimensional subspace spanned by $|\pm\alpha\rangle$~\cite{Mirrahimi2014, Leghtas2015}.
Bit-flips between these $Z$ logical states are exponentially suppressed in $|\alpha|^2$, but phase-flips are not.
As shown in Fig.~\ref{fig:gate_concept}(a), any $X$ rotation whose trajectory stays on the Bloch sphere surface must cross the equator, whose states (e.g., $|\mathcal{C}_\pm\rangle$ on the $X$ axis and $|\alpha\rangle\pm i|{-}\alpha\rangle$ on the $Y$ axis) are not error-protected.
Any bias-preserving $X$ gate must therefore leave the two-dimensional code space, taking a path through the larger Hilbert space of the oscillator that connects $|\alpha\rangle$ to $|{-}\alpha\rangle$~\cite{Guillaud2022lectures, Puri2020}.
For a single target cat ($\hat{a}_t$), this can be achieved by applying a frequency-shift Hamiltonian $\lambda \hat{a}_t^\dagger \hat{a}_t$ that rotates the coherent state along a circular path $\alpha e^{i\theta}$ through phase space [Fig.~\ref{fig:gate_concept}(b)]~\cite{Mirrahimi2014, Puri2020}.
To promote this to a CNOT between control and target cats, the frequency shift $\lambda$ must be conditioned on the control's logical states $\approx|\pm\alpha_c\rangle$, which are approximate eigenstates of the non-Hermitian operator $\hat{a}_c$ (control cat mode).
The Hermitian realization is a longitudinal coupling $\hat{H}_{\rm int} = \lambda(\hat{a}_c^\dagger + \hat{a}_c)\hat{a}_t^\dagger \hat{a}_t$~\cite{Guillaud2019, Gautier2023zeno}, but because coherent states are not exact eigenstates of the position operator $(\hat{a}_c^\dagger + \hat{a}_c)$, this leads to target bit-flip errors and control phase-flip errors from displacement backaction.
These errors are typically suppressed by keeping the two-photon stabilization $\kappa_2$ active on the control cat during the gate, using the quantum Zeno effect to confine the dynamics~\cite{Guillaud2019, Gautier2023zeno}.
However, when optimized jointly against single-photon loss and gate time, the residual phase-flip error scales only as ${\sim}1/\sqrt{\kappa_2}$, limiting the control phase-flip to the percent level in practice~\cite{Guillaud2021, Chamberland2022}.

We propose a fundamentally different approach: Instead of a frequency shift, we SWAP the target cat into an auxiliary ``bin'' mode and back, which accumulates a geometric phase that effects the $X$ gate [Fig.~\ref{fig:gate_concept}(c)].
We emphasize that ``SWAP'' here refers to the physical exchange of photon populations between two bosonic modes via a beam-splitter interaction~\cite{deGraaf2025, YurkeMcCallKlauder1986, CamposSalehTeich1989}, not a logical-state swap gate.  On the single-photon subspace it coincides with the familiar two-qubit iSWAP; on the multi-photon coherent-state input relevant here it acquires a per-photon $-i$ phase that composes to $(-1)^{\hat n}$ over two SWAPs, so SWAP$^2$ is not the identity and the $|\alpha\rangle \to |{-}\alpha\rangle$ mapping realises the cat $X$ gate (Supplemental Material~\cite{SM}, Sec.~\ref{sm-sec:sm-su2}).
At first glance, routing the target cat through the bin and back via SWAP$^2$ appears to destroy the target's bit-flip protection.
However, the beam-splitter unitarily transfers the coherent state (and its exponential bit-flip suppression) intact to the bin; the protection resides in whichever mode holds the state throughout the SWAP$^2$ cycle (Sec.~\ref{sm-sec:sm-biasswap} of~\cite{SM}).

To promote this SWAP-based $X$ gate to a CNOT, we add a cross-Kerr nonlinear coupling between the control cat and the bin, so the full gate involves three modes, control cat ($\hat{a}_c$), auxiliary bin ($\hat{b}$), and target cat ($\hat{a}_t$), governed by [Fig.~\ref{fig:gate_concept}(d)]
\begin{equation}\label{eq:hamiltonian}
\hat{H} = \chi\, \hat{a}_c^\dagger \hat{a}_c\, \hat{b}^\dagger \hat{b} \;+\; g\!\left(\hat{b}^\dagger \hat{a}_t + \hat{b}\,\hat{a}_t^\dagger\right),
\end{equation}
where the first term is a cross-Kerr interaction between control and bin~\cite{Holland2015, Ye2021quarton, Arne}, and the second is a beam-splitter coupling between bin and target~\cite{Zhang2019beamsplitter, Gao2019}.
The bin mode starts and ends in vacuum and serves only as a mediator during the gate; nevertheless, during the mid-cycle of SWAP$^2$ the target cat resides in the bin, so the bin's single-photon lifetime and self-Kerr requirements are comparable to those of the target cavity.
(A parametric longitudinal coupling $\lambda\,(\hat{a}_c + \hat{a}_c^\dagger)\,\hat{b}^\dagger\hat{b}$ can replace the cross-Kerr term as an alternative VCB implementation; see Sec.~\ref{sm-sec:sm-longitudinal} of~\cite{SM}.)

The CNOT's conditional activation mechanism is closely analogous to the CR$_X$ gate in cat-transmon architectures~\cite{Hann2024}.
The unconditional displacement $\hat{D}(\alpha_c)$ maps the control cat from the $\{|\alpha_c\rangle, |{-}\alpha_c\rangle\}$ basis to $\{|2\alpha_c\rangle, |0\rangle\}$, respectively [Fig.~\ref{fig:gate_concept}(e)].
When the control is in $|0\rangle$ (ON), the cross-Kerr term vanishes and the beam-splitter drive is resonant.
It then executes a full SWAP$^2$ cycle (two elementary SWAPs, each of duration $T_{\rm SWAP} = \pi/(2g)$, so that the target state transfers entirely into the bin and returns), accumulating a geometric phase that maps $|\alpha_t\rangle \to |{-}\alpha_t\rangle$, i.e., the $X$ gate (see Sec.~\ref{sm-sec:sm-su2} of~\cite{SM}).
When the control is in $|2\alpha_c\rangle$ (OFF), the cross-Kerr interaction shifts the bin frequency by $\chi n_c$ per control Fock component, with average detuning $\bar{\Delta} = \chi\langle\hat{n}_c\rangle = 4\chi|\alpha_c|^2$, making the beam-splitter drive far off-resonant and leaving the target unaffected.
This vacuum-conditional selectivity implements a CNOT in the cat qubit basis.
After the gate, the control is displaced back to the $\{|\alpha_c\rangle, |{-}\alpha_c\rangle\}$ basis; in practice, this final displacement is implemented as part of a displacement echo that also suppresses several control error channels, as discussed below.

Crucially, the two-photon stabilization $\kappa_2$ on both cats is turned off during the gate and restored afterwards.
The entire operation is coherent, which alleviates the gate speed limits imposed by schemes that require dissipation to remain active~\cite{Guillaud2019, Gautier2023zeno}, and enables gate times set by the interaction strengths $\chi$ and $g$ rather than the dissipation rate $\kappa_2$.
Recent experiments have shown that intermittent stabilization (turning $\kappa_2$ off and back on) does not significantly degrade the cat qubit's noise bias~\cite{Putterman2024cat, Reglade2024}.

\begin{figure}[t]
\includegraphics[width=\columnwidth]{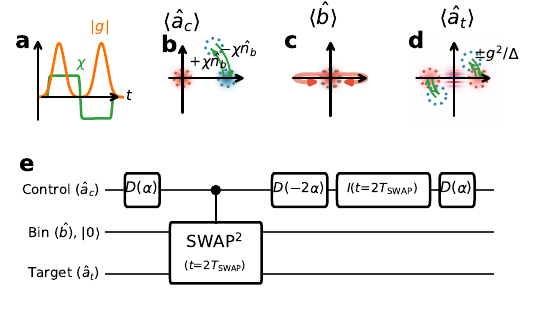}
\caption{\label{fig:chiecho}
Echo techniques for gate error suppression.
(a)~$\chi$-echo pulse sequence: The cross-Kerr coupling $\chi$ (green) flips sign at the gate midpoint while two $\pi/2$ beam-splitter pulses $|g|$ (orange) execute the conditional SWAP$^2$.
(b)--(d)~Phase-space diagrams for the control, bin, and target modes.
In~(b), the OFF-state control $|2\alpha_c\rangle$ acquires a cross-Kerr phase $\pm\chi\hat{n}_b$ that is refocused by the $\chi$-echo.
In~(c), the $\chi$-echo is invisible to the ON branch (control in vacuum has no cross-Kerr shift); the bin still returns to vacuum after the complete SWAP$^2$ cycle.
In~(d), the target cat acquires a $g^2/\bar{\Delta}$ AC~Stark phase in the OFF branch, which is also refocused by the $\chi$-echo.
(e)~Full gate protocol with displacement echo: The control is displaced by $\hat{D}(\alpha_c) = \exp(\alpha_c\hat{a}_c^\dagger - \alpha_c^*\hat{a}_c)$, a conditional SWAP$^2$ acts on the bin--target pair, then the displacement echo $\hat{D}(-2\alpha_c)$--idle--$\hat{D}(\alpha_c)$ balances phases from e.g.\ self-Kerr or dephasing between the two logical branches.
}
\end{figure}

\textit{Gate error suppression via echoes}---We employ two echo techniques, a $\chi$-echo and a control-cat displacement echo~\cite{Hann2024}, to suppress coherent gate errors.  The required echo displacements $\hat{D}(\alpha_c)$ and $\hat{D}(-2\alpha_c)$ are unconditional coherent drives on the control cat whose duration (a few nanoseconds at typical drive strengths) is negligible compared to $T_{\rm SWAP}$ and $T_{\rm meas}$; we treat them as instantaneous throughout.

In the OFF branch, the beam-splitter drive is detuned by $\bar{\Delta} = 4\chi|\alpha_c|^2$, which suppresses bin population to a small but non-zero transient $\sim(g/\bar{\Delta})^2$ during the pulse [Fig.~\ref{fig:chiecho}(c)].
Realistically, the bin also carries a residual thermal population $\bar{n}_{\rm th}$.
Through the cross-Kerr interaction, both the coherent transient and the thermal bin population imprint a photon-number-dependent phase $e^{i\chi\hat{n}_b t}$ on the displaced control state $|2\alpha_c\rangle$ [Fig.~\ref{fig:chiecho}(b)].
Without correction, this would produce both a large control phase-flip and a control bit-flip.
A $\chi$-echo protocol, which flips the sign of the cross-Kerr at the gate midpoint [Fig.~\ref{fig:chiecho}(a); see Sec.~\ref{sm-sec:sm-chiecho} of~\cite{SM} for details], refocuses this accumulated phase, canceling the leading-order $\chi T_{\rm SWAP}$ rotation [Fig.~\ref{fig:chiecho}(b)].
The off-resonant beam-splitter drive also imparts a deterministic AC Stark shift $\sim g^2/\bar{\Delta}$ on both the target and bin modes [Fig.~\ref{fig:chiecho}(d)].
On the target, this rotates the cat-state axis; upon subsequent stabilization, the rotated axis reveals whether SWAP$^2$ occurred, effectively leaking which-path information about the control state and causing a control phase-flip.
The $\chi$-echo changes the sign of $\bar{\Delta}$, which cancels this AC Stark phase to leading order (Sec.~\ref{sm-sec:sm-beamsplitter-detuned} of~\cite{SM}).
Alternatively, if the $\chi$ sign flip is difficult to implement, the control phase from bin $\bar{n}_{\rm th}$ can be nulled by choosing a gate time such that the control $|2\alpha_c\rangle$ completes full rotation cycles (Sec.~\ref{sm-sec:sm-cycling-condition} of~\cite{SM}), while the target AC Stark phase can be corrected by a small detuning of the second cSWAP pulse (Sec.~\ref{sm-sec:sm-beamsplitter-detuned} of~\cite{SM}).

In the displaced basis $\{|0\rangle, |2\alpha_c\rangle\}$, any photon-number-preserving channel on the control, such as the parasitic self-Kerr $K_a \hat{a}^{\dagger 2}\hat{a}^2/2$ or Markovian dephasing $\mathcal{D}[\hat{a}^\dagger\hat{a}]$, is no longer parity-preserving and thus induces phase-flip errors scaling as $(K_a T_{\rm SWAP})^2$ for the coherent Kerr contribution; see Sec.~\ref{sm-sec:sm-dispecho} and Sec.~\ref{sm-sec:sm-displaced-dephasing} of~\cite{SM} for Kerr and dephasing scaling, respectively.
A displacement echo [Fig.~\ref{fig:chiecho}(e)], first proposed for CR$_X$ gates~\cite{Hann2024}, balances these phases between the two logical branches by displacing the control from $\{|0\rangle, |2\alpha_c\rangle\}$ to $\{|{-}2\alpha_c\rangle, |0\rangle\}$ at the gate midpoint, so that each branch accumulates the same phase over half the gate.  This converts the scaling from $(K_a T_{\rm SWAP})^2$ to $(K_a T_{\rm SWAP})^4$, suppressing the control phase-flip from self-Kerr by several orders of magnitude, at the cost of doubling the total gate time to $T_{\rm CX} = 4\,T_{\rm SWAP}$.  The second half of the displacement echo is an idle period (no beam-splitter drive); the control simply evolves under its self-Kerr for $T_{\rm SWAP}$, accumulating the same nonlinear phase as the first half.  For the echo to cancel, the control must experience the same self-Kerr $K_a$ in both halves of the protocol (Sec.~\ref{sm-sec:sm-dispecho} of~\cite{SM}; an idle-free variant that avoids this overhead is discussed in Sec.~\ref{sm-sec:sm-idle-free-echo} of~\cite{SM}).

\begin{figure*}[t]
\includegraphics[width=\textwidth]{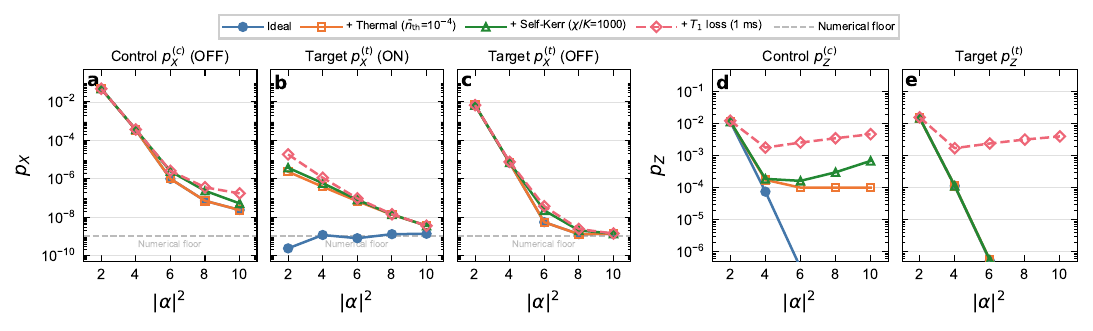}
\caption{\label{fig:buildup}
VCB gate performance.
Each panel tracks one error channel versus $|\alpha|^2$ ($= |\alpha_c|^2 = |\alpha_t|^2$): (a) control OFF bit-flip, (b) target ON bit-flip, (c) target OFF bit-flip, (d) control phase-flip, and (e) target phase-flip.
Successive curves cumulatively add imperfections.
\textbf{Blue}: Ideal Hamiltonian (cross-Kerr + beam-splitter + $\chi$-echo + DRAG, vacuum initial states).
\textbf{Orange}: Adds thermal population ($\bar{n}_{\rm th} = 10^{-4}$) to both control and bin (negligible for target).
\textbf{Green}: Adds self-Kerr on all three modes ($K_a = K_b = K_t = \chi/1000$) with displacement echo on the control phase-flip channel.
\textbf{Red dashed}: Adds single-photon loss ($T_1 = 1$\,ms) analytically---target phase-flip via $(1 - e^{-2\bar{n}T_{\rm CX}/T_1})/2$, control phase-flip via $(1 - e^{-2\bar{n}T_{\rm CX}/T_1})/2$ (same rate as the undisplaced cat, since in the Heisenberg picture the identity component of $\hat{a}+\alpha$ does not induce phase flips), bit-flip via $\bar{n}\,e^{-2\bar{n}}\,T_{\rm CX}/T_1$, plus a bin thermal heating floor $2\bar{n}_{\rm th}\,T_{\rm CX}/T_1$ on the control bit-flip (validated in Fig.~\ref{sm-fig:control_bin_T1} of~\cite{SM}).
Grey dashed line: Numerical floor (${\approx}\,10^{-9}$; see Sec.~\ref{sm-sec:sm-simulations} of~\cite{SM}).
Parameters: Elementary SWAP duration $T_{\rm SWAP} = 100$\,ns (Sec.~\ref{sm-sec:swap-glossary}), $\chi/2\pi = 4$\,MHz, $\sigma = 0.18$ (see Sec.~\ref{sm-sec:sm-drag} of~\cite{SM} for pulse-shaping details).
Total protocol time with displacement echo: $T_{\rm CX} = 4\,T_{\rm SWAP} = 400$\,ns.
}
\end{figure*}

\textit{Gate performance}---We simulate the full three-mode CNOT gate unitarily. The large Hilbert space dimension at $\bar{n} \approx 10$ makes Lindblad master-equation simulation prohibitively expensive, so we account for dissipative channels---photon loss and thermal heating---analytically, assuming $T_1 = 1$~ms throughout and validated against Lindblad simulations of the subsystems (Sec.~\ref{sm-sec:T1_comparison}).
We include both $\chi$-echo and control displacement echo [Fig.~\ref{fig:chiecho}(e)], tracking five error channels: target bit-flips (ON and OFF), control bit-flip (OFF), and control and target phase-flips.  We denote the bit-flip and phase-flip error probabilities as $p_X^{(j)}$ and $p_Z^{(j)}$ ($j \in \{c, t\}$ for control and target). Formal definitions via projection onto the cat-qubit basis are given in Sec.~\ref{sm-sec:sm-error-defs} of~\cite{SM}.  We parametrize the cat size by $\bar{n} = |\alpha_c|^2 = |\alpha_t|^2$.
The control ON bit-flip is identically zero since the control is in vacuum, an eigenstate of the gate Hamiltonian ($\hat{n}_c|0\rangle = 0$).
Figure~\ref{fig:buildup} reveals which imperfection dominates each channel as we progressively add thermal population, self-Kerr, and photon loss.
Here we assume a residual thermal population $\bar{n}_{\rm th}=10^{-4}$ (Sec.~\ref{sm-sec:thermal_sensitivity} of~\cite{SM}).
The control OFF bit-flip [Fig.~\ref{fig:buildup}(a)] is the hardest to suppress: It is limited by both the gate speed, which sets the adiabaticity ratio $g_{\max}/\bar{\Delta}$ (Sec.~\ref{sm-sec:sm-drag} of~\cite{SM}), and bin thermal dissipation, which imparts an irreversible cross-Kerr rotation on the control (Sec.~\ref{sm-sec:T1_comparison} of~\cite{SM}).
Shaping the envelope $g(t)$ as a Gaussian pulse suppresses the coherent leakage exponentially with the pulse width~$\sigma$, and an additional DRAG derivative correction~\cite{Motzoi2009} reduces the residual by ${\sim}3\times$ (Sec.~\ref{sm-sec:sm-drag} of~\cite{SM}).
The target ON bit-flip [Fig.~\ref{fig:buildup}(b)] would be zero for a perfectly resonant SWAP$^2$, but the assumed residual thermal population $\bar{n}_{\rm th}=10^{-4}$ in the control ON ($|0\rangle$) state introduces a small probability of the beam-splitter being detuned by $\Delta = \chi$, preventing the SWAP from completing perfectly (Sec.~\ref{sm-sec:sm-thermal-target-on} of~\cite{SM}).
The target OFF bit-flip [Fig.~\ref{fig:buildup}(c)] is not quite coherence-limited at low $\bar{n}$ because the detuning $\bar{\Delta} = 4\chi|\alpha_c|^2$ is insufficient to fully suppress the beam-splitter; at $\bar{n} \gtrsim 6$ the detuning is large enough that this channel is well below the coherence limit.
Both phase-flip channels [Fig.~\ref{fig:buildup}(d)--(e)] are well below the coherence limit $\bar{n}\,T_{\rm CX}/T_1$, where $T_{\rm CX} = 4\,T_{\rm SWAP}$ is the total gate duration including the displacement echo.
The target phase-flip [Fig.~\ref{fig:buildup}(e)] is especially well protected: The beam-splitter between bin and target conserves total photon number and thus preserves the target cat parity, so intrinsic gate-induced phase-flips are negligible.

The control phase-flip [Fig.~\ref{fig:buildup}(d)] has both coherent and dissipative contributions.
The coherent contribution has a local source, control self-Kerr distortion, which is suppressed from $(K_a T_{\rm SWAP})^2$ to $(K_a T_{\rm SWAP})^4$ by the displacement echo (Sec.~\ref{sm-sec:sm-dispecho} of~\cite{SM}).
The coherent contribution also has non-local sources: bin thermal population ($p_Z^{(c)} \approx \bar{n}_{\rm th}$, irreducible) and bin--target Kerr mismatch (Sec.~\ref{sm-sec:sm-ctrl-pf} of~\cite{SM}).
After all echoes, the total saturates at ${\sim}10^{-4}$, assuming approximate Kerr-matching ($K_b \approx K_t$) and $\chi/K = 10^{3}$ (i.e.\ $\chi/2\pi\approx 4$~MHz with $K/2\pi\approx 4$~kHz); see Sec.~\ref{sm-sec:sm-kerr-mismatch} of~\cite{SM} for the Kerr-mismatch sensitivity.
If the realized bin occupation is instead $\bar{n}_{\rm th}=10^{-3}$ or $5\times10^{-3}$, this irreducible thermal phase-flip floor increases linearly to the same order.
Photon loss with timescale $T_1$ adds linearly to the phase-flip budget.
Both control and target contribute $\bar{n}\, T_{\rm CX}/T_1$: Although the displaced control state $|2\alpha_c\rangle$ has mean photon number $4\bar{n}$, in the Heisenberg picture $\hat{a}_c \to \hat{a}_c + \alpha_c = \alpha_c(\hat{Z} + \hat{I})$ (projected into the codespace). The identity component does not induce phase flips, leaving the dephasing rate at $\kappa_1\bar{n}$.
At $\bar{n} = 10$ and $T_1 = 1$\,ms, this is ${\approx}\,4\times 10^{-3}$ per gate for each cat.

\begin{table}[b]
\caption{\label{tab:error_budget}Gate error budget at $\bar{n} = 10$, $T_1 = 1$\,ms, elementary SWAP duration $T_{\rm SWAP} = 100$\,ns (see Fig.~\ref{fig:buildup}).  Coherent and dissipative mechanisms are separated where relevant.}
\vspace{16pt}
\begin{ruledtabular}
\begin{tabular}{lcr}
Channel & Dominant mechanism & Value \\
\hline
\multirow{2}{*}{Control $p_X^{(c)}$ (OFF)} & Adiabaticity + self-Kerr & $5.2 \times 10^{-8}$ \\
 & Bin thermal + $T_1$ & $1.7 \times 10^{-7}$ \\
Target $p_X^{(t)}$ (ON) & Thermal $\bar{n}_{\rm th}$ in control & $3.5 \times 10^{-9}$ \\
Target $p_X^{(t)}$ (OFF) & Negligible (no SWAP) & $1.4 \times 10^{-9}$ \\
\hline
\multirow{2}{*}{Control $p_Z^{(c)}$} & Thermal + self-Kerr & $6.8 \times 10^{-4}$ \\
 & Photon loss & $4.0 \times 10^{-3}$ \\
\hline
\multirow{2}{*}{Target $p_Z^{(t)}$} & Negligible (parity conserved) & $2.7 \times 10^{-8}$ \\
 & Photon loss & $4.0 \times 10^{-3}$ \\
\end{tabular}
\end{ruledtabular}
\end{table}

Table~\ref{tab:error_budget} summarizes the error budget at $\bar{n} = 10$.
All bit-flip channels remain exponentially suppressed (${\lesssim}10^{-7}$ for control, ${\lesssim}10^{-9}$ for target), preserving the noise bias, while coherent (pre-loss) phase-flip errors saturate below $10^{-3}$ for the assumed thermal population used in Fig.~\ref{fig:buildup}; the total per-gate phase-flip $\sim\!4\times10^{-3}$ is dominated by $T_1$.  The gate is coherence-limited: Improving $T_1$ and thermalization directly reduces the residual errors.
Decreasing the gate time further reduces the coherence-limited errors, but at the cost of increased bit-flip errors from the stronger drive.  We use $T_{\rm SWAP} = 100$~ns here as a general operating point suited to transversal logical CNOTs (where the control OFF bit-flip must also be exponentially suppressed); in the next section we switch to $T_{\rm SWAP} = 50$~ns, which is optimal for syndrome extraction.  A systematic comparison of this trade-off across $T_{\rm SWAP} = 25$--$100$~ns is given in Sec.~\ref{sm-sec:sm-gate-time} of~\cite{SM}.

\begin{figure}[t]
\includegraphics[width=\columnwidth]{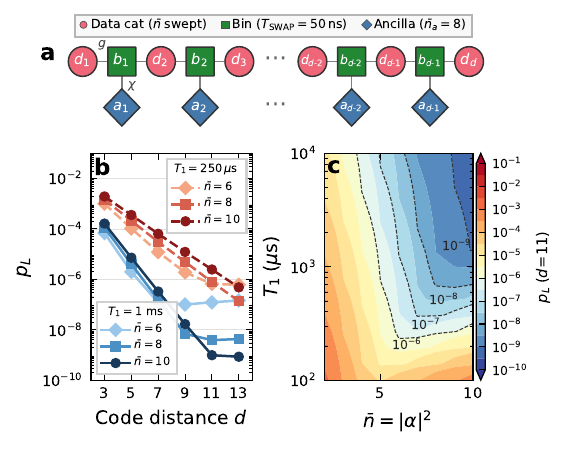}
\caption{\label{fig:performance}
Repetition code performance with the VCB gate (elementary SWAP duration $T_{\rm SWAP} = 50$\,ns).
(a)~Physical layout: data cats (green), ancilla cats (blue), and a single shared bin mode (pink diamond) per ancilla in a linear chain.
(b)~Logical error per syndrome cycle versus code distance $d$ for $\bar{n} = 6$, $8$, and $10$ at $T_1 = 1$~ms (solid) and $T_1 = 250\,\mu$s (dashed).
(c)~Logical error per cycle at $d=11$ as a function of $\bar{n}$ and $T_1$.
Parameters: $\chi/2\pi = 4$~MHz, $\sigma = 0.22$, $\bar{n}_{\rm th} = 10^{-4}$, $\bar{n}_a = 8$. See Sec.~\ref{sm-sec:sm-drag} of~\cite{SM} for pulse-shaping details.
}
\end{figure}

\textit{Repetition code performance}---We simulate the logical error rate in a distance-$d$ repetition cat code~\cite{Guillaud2019, Guillaud2021, Chamberland2022} [Fig.~\ref{fig:performance}(a)] using a parallel syndrome-extraction schedule (Sec.~\ref{sm-sec:sm-simcnot} of~\cite{SM}) with an effective gate time $T_{\rm CX} = 4\,T_{\rm SWAP} = 200$~ns at elementary SWAP duration $T_{\rm SWAP} = 50$~ns.
Including a measurement time $T_{\rm meas} = 200$~ns (which subsumes ancilla measurement and reset/re-stabilisation; Sec.~\ref{sm-sec:sm-simcnot} of~\cite{SM}), the total syndrome-extraction round time is $2\,T_{\rm CX} + T_{\rm meas} = 600$~ns.

We emphasize that the ancilla remains in the displaced basis $\{|0\rangle, |2\alpha_c\rangle\}$ throughout both CNOT steps, with the displacement echo and restabilization applied only after all gates in the round are complete.  This allows us to exploit the unitary character of the vacuum-conditional mechanism to protect the data cat from ancilla errors.  In a standard syndrome circuit, an ancilla bit-flip between an ancilla's two CNOTs propagates as an $X$ error onto the second data cat; in existing cat--cat CNOT gates with always-on ancilla stabilization, such bit-flips are realized continuously during the gate~\cite{Guillaud2019, Gautier2023zeno}.  Here, in contrast, ancilla bit-flips are dominated by $|2\alpha_c\rangle$ phase-space rotations (Fig.~\ref{fig:chiecho}(b)) that only manifest at the end of syndrome extraction, when the control is displaced back to the $\{|\alpha_c\rangle, |{-}\alpha_c\rangle\}$ basis and re-stabilized; at that point the only requirement is the easy one that $p_X^{(c)} \not\gg p_Z^{(c)}$, because these bit-flips are reset before they can propagate to the data cat (Sec.~\ref{sm-sec:sm-simcnot} of~\cite{SM}).
The full noise model is detailed in Sec.~\ref{sm-sec:sm-simcnot} of~\cite{SM}; the per-gate error rates used as inputs are shown in Fig.~\ref{sm-fig:qec_gate_input} of~\cite{SM}.
Here we use the same assumed residual thermal population $\bar{n}_{\rm th}=10^{-4}$ as in Fig.~\ref{fig:buildup} (Sec.~\ref{sm-sec:sm-ctrl-pf-thermal} of~\cite{SM}).

Figure~\ref{fig:performance}(b) shows the logical error per syndrome cycle $p_L$ at $T_1 = 1$~ms~\cite{Reagor2016, Acharya2026TaResonators} and $T_1 = 250\,\mu$s.
At $T_1 = 1$~ms and $\bar{n}=10$, the logical error drops below $10^{-6}$ at $d=7$ (13 cats), and reaches ${\sim}10^{-9}$ at $d=11$ (21 cats).
Reducing $T_1$ to $250\,\mu$s raises the logical error to ${\sim}10^{-6}$ at $d=11$---a factor-of-${\sim}10^3$ degradation, highlighting cavity coherence as a main hardware requirement.
A comparison using the same repetition-code simulation setup with an optimized Zeno CNOT is given in Sec.~\ref{sm-sec:sm-zeno-qec-comparison} of~\cite{SM}.
Figure~\ref{fig:performance}(c) maps the achievable logical error at fixed $d=11$ across the $(\bar{n},\,T_1)$ parameter space.  Increasing the cat size $\bar{n}$ becomes increasingly beneficial at longer $T_1$: The exponential bit-flip suppression from larger $\bar{n}$ can only be exploited once the phase-flip rate $\sim\bar{n}/T_1$ is sufficiently low.

\textit{Discussion and outlook}---We have proposed a bias-preserving CNOT gate between dissipative cat qubits that operates entirely via coherent dynamics: A vacuum-conditional beam-splitter mediated by cross-Kerr coupling requires no engineered dissipation during the gate.
The gate preserves exponential bit-flip suppression for both control and target cats while remaining coherence-limited in phase-flip.
Crucially, the VCB gate avoids the non-adiabatic error channel associated with Zeno-based gates~\cite{Guillaud2019, Gautier2023zeno}, removing one of the key constraints in existing repetition-cat architectures~\cite{Gouzien2023}.
When used in syndrome extraction, this gate additionally shields the data cat from nearly all ancilla errors: The data cat's conditional $X$ only requires the ancilla to provide a coarse ``$n_c$ near zero versus $n_c \gg 0$'' state separation while dominant sources of ancilla bit-flips are phase-space rotations of the $|2\alpha_c\rangle$ state that only map to bit-flips during stabilization after the unitary gate evolution.  Together these properties enable logical memory with $p_L < 10^{-6}$ per cycle in a distance-$7$ repetition code consisting of 13 cat qubits.

Because the gate preserves noise bias for both control and target cats, it can also be used transversally between repetition-code blocks, which---combined with recent work on algorithmic fault tolerance~\cite{Zhou2024algFT, Cain2024correlated}---can speed up logical computation.
The scheme extends naturally to a bias-preserving Toffoli gate (Sec.~\ref{sm-sec:sm-toffoli} of~\cite{SM}), which reduces the overhead of magic-state distillation.
Combined with LDPC-cat codes that offer ${\sim}5\times$ higher encoding rates than repetition codes~\cite{Ruiz2024}, the VCB gate set provides a path toward reduced-overhead fault-tolerant quantum computation; a detailed resource estimate is an important direction for future work.

The proposed gate is experimentally realizable with circuit-QED ingredients.  The beam-splitter component is comparatively mature: strong, high-on/off exchange interactions between bosonic modes have been demonstrated~\cite{Lu2023, Chapman2023}.  The remaining nonlinear ingredient is a large cross-Kerr with suppressed self-Kerr; this direction has seen substantial experimental progress in single-photon-resolved cross-Kerr interactions~\cite{Holland2015} and recent quarton-coupler devices~\cite{Ye2025quarton}.  Details and a possible circuit path are discussed in Sec.~\ref{sm-sec:sm-hardware-realizations} of~\cite{SM}.

We thank Matt Matheny, Akshay Koottandavida, Hanho Lee, Shahriar Aghaei, Fernando Brandao, Maximilian Seifert, and Andrew Keller for fruitful discussions and reviews.
This work was completed while K.N. was a researcher at Amazon Center for Quantum Computing.
A patent application related to this work has been filed.
\putbib[main]
\end{bibunit}
\makeatletter
\global\let\@FMN@list\@empty
\makeatother

\clearpage
\onecolumngrid

\begin{center}
\rule{\textwidth}{0.5pt}\\[1em]
{\Large \textbf{Supplemental Material}}\\[0.5em]
\rule{\textwidth}{0.5pt}
\end{center}

\vspace{2em}

\begin{enumerate}[label=\textbf{S-\Roman*.}, leftmargin=2.5em, itemsep=0.4em]
\item \hyperref[sec:sm-error-defs]{Error definitions and notation}
\item \hyperref[sec:sm-beamsplitter]{Beam-splitter general solution (resonant and detuned)}
\item \hyperref[sec:sm-su2]{SU(2) equivalence, geometric phase, and CR$_X$ analogy}
\item \hyperref[sec:sm-biasswap]{Target cat bias preservation during SWAP}
\item \hyperref[sec:sm-control]{Control cat bias preservation with cross-Kerr coupling to bin}
\item \hyperref[sec:sm-chiecho]{$\chi$-echo: more details}
\item \hyperref[sec:sm-dispecho]{Control phase-flip: local sources and displacement echo}
\item \hyperref[sec:sm-ctrl-pf]{Control phase-flip: non-local sources and Kerr-matching}
\item \hyperref[sec:sm-idle-free-echo]{Idle-free control displacement echo}
\item \hyperref[sec:sm-longitudinal]{Longitudinal-coupling variant of the VCB mechanism}
\item \hyperref[sec:sm-hardware-realizations]{Potential hardware realizations}
\item \hyperref[sec:sm-drag]{Pulse shaping}
\item \hyperref[sec:sm-simcnot]{Repetition code QEC: more details}
\item \hyperref[sec:sm-simulations]{Simulation methods and convergence}
\item \hyperref[sec:sm-toffoli]{Toffoli gate extension}
\end{enumerate}

\clearpage
\twocolumngrid

\renewcommand{\thepage}{S\arabic{page}}
\setcounter{page}{1}
\setcounter{secnumdepth}{3}
\renewcommand{\thesection}{S-\Roman{section}}
\setcounter{section}{0}
\renewcommand{\thetable}{S\arabic{table}}
\setcounter{table}{0}
\renewcommand{\theequation}{S\arabic{equation}}
\setcounter{equation}{0}
\renewcommand{\thefigure}{S\arabic{figure}}
\setcounter{figure}{0}
\renewcommand{\theHfigure}{S\arabic{figure}}
\renewcommand{\theHtable}{S\arabic{table}}
\renewcommand{\theHequation}{S\arabic{equation}}
\renewcommand{\theHsection}{S-\Roman{section}}
\renewcommand{\theHsubsection}{S-\Roman{section}.\Alph{subsection}}
\renewcommand{\theHsubsubsection}{S-\Roman{section}.\Alph{subsection}.\arabic{subsubsection}}

%
%
\begingroup
\makeatletter
\def\@extra@b@citeb{-sm}%
\def\@extra@binfo{-sm}%
\begin{bibunit}[apsrev4-2]

\section{Error definitions and notation}
\label{sec:sm-error-defs}
In this section we define the cat-qubit basis states, error extraction
procedure, and notation conventions used throughout the paper and
supplement.

\subsection{Cat-qubit basis and stabilization}
\label{sec:cat-basis}

The cat qubit is encoded in the two-dimensional subspace spanned by
the even- and odd-parity coherent-state superpositions
\begin{equation}\label{eq:cat_states}
|\mathcal{C}_\pm\rangle = \mathcal{N}_\pm\bigl(|\alpha\rangle \pm |{-}\alpha\rangle\bigr),
\end{equation}
where $\mathcal{N}_\pm = (2 \pm 2e^{-2|\alpha|^2})^{-1/2}$ are
normalization constants.  The logical states are the approximate
coherent states $|0_L\rangle \approx |\alpha\rangle$ and
$|1_L\rangle \approx |{-}\alpha\rangle$ (exact in the limit
$|\alpha|^2 \to \infty$), which are eigenstates of the Pauli $Z$
operator in the cat basis.

Engineered two-photon dissipation
$\dot{\rho} = \kappa_2\,\mathcal{D}[\hat{a}^2 - \alpha^2]\rho$
confines the oscillator to the cat-qubit subspace on a timescale
$1/\kappa_2$~\cite{Mirrahimi2014, Leghtas2015}.

\subsection{Error extraction via projection}
\label{sec:error-extraction}

Throughout this work, gate errors are quantified by applying the gate
(with stabilization \emph{off}) and then projecting the output state
onto the cat-qubit basis.  This projection is equivalent to applying
infinite-time idealized stabilization
$\mathcal{L}_0 = \kappa_2\,\mathcal{D}[\hat{a}^2 - \alpha^2]$ and
reading off the resulting logical state.
We emphasize that the two-photon stabilization is turned off during
the gate and restored only afterwards; the projection at the end
models this restoration.  Infinite-time stabilization removes leakage
from the cat manifold; it does not correct logical bit-flip or
phase-flip errors within the manifold.

Concretely, after evolving the full three-mode state
$\rho(T_{\mathrm{gate}})$ under the gate Hamiltonian, the control and target reduced density
matrices $\rho_c$ and $\rho_t$ are extracted by partial trace.  For the
control (which operates in the displaced basis during the gate), the displacement $\hat{D}(-\alpha_c)$ is
applied to return to the cat frame before projection.

\subsection{Bit-flip and phase-flip error definitions}
\label{sec:error-defs}

For a finite-time gate channel $\mathcal{E}$ acting on a state $\rho$
in the cat manifold, followed by infinite-time idealized stabilization,
\begin{equation}\label{eq:rho-inf}
\begin{aligned}
\rho_\infty
&= \lim_{t\to\infty} e^{\mathcal{L}_0 t}\,\mathcal{E}(\rho) \\
&= c_{++}|\mathcal{C}_+\rangle\langle\mathcal{C}_+|
 + c_{--}|\mathcal{C}_-\rangle\langle\mathcal{C}_-| \\
&\quad
 + c_{+-}|\mathcal{C}_+\rangle\langle\mathcal{C}_-|
 + c_{+-}^*|\mathcal{C}_-\rangle\langle\mathcal{C}_+| .
\end{aligned}
\end{equation}
The coefficients are determined by the conserved quantities
$\hat{J}_{++}$ and $\hat{J}_{+-}$ of the two-photon stabilizer
Lindbladian~\cite{Mirrahimi2014}: $c_{++} =
\mathrm{Tr}[\hat{J}_{++}^{\dagger}\mathcal{E}(\rho)]$ and
$c_{+-} = \mathrm{Tr}[\hat{J}_{+-}^{\dagger}\mathcal{E}(\rho)]$.
Here $\hat{J}_{++}$ is the even-parity conserved quantity and
$\hat{J}_{+-}$ is the even--odd coherence conserved quantity.

We define the bit-flip probability $p_X$ as the probability that
$|0_L\rangle \to |1_L\rangle$ after $\mathcal{E}$ and stabilization,
and the phase-flip probability $p_Z$ as the probability that
$|\mathcal{C}_+\rangle \to |\mathcal{C}_-\rangle$.  Thus,
\begin{align}
\label{eq:bitflip_def}
p_X &= \frac{1}{2}
- \mathrm{Re}\!\left(\mathrm{Tr}\!\left[
\hat{J}_{+-}\,\mathcal{E}(|0_L\rangle\langle 0_L|)
\right]\right),\\
\label{eq:phaseflip_def}
p_Z &= 1 - \mathrm{Tr}\!\left[
\hat{J}_{++}\,\mathcal{E}(|\mathcal{C}_+\rangle\langle\mathcal{C}_+|)
\right].
\end{align}
The main text labels these quantities $p_X^{(j)}$ and $p_Z^{(j)}$ for
mode $j\in\{c,t\}$.

\subsection{``Idle'' evolution}
\label{sec:idle-def}

Several sections compare gate errors against a baseline of ``idle''
evolution.  By this we mean free Hamiltonian evolution (e.g., under
self-Kerr $(K/2)\,\hat{a}^{\dagger 2}\hat{a}^2$) \emph{without}
two-photon stabilization active, followed by projection onto the cat
basis to extract errors as defined above.  This models the scenario
where stabilization is off during the gate duration and restored
afterwards.

\subsection{Beam-splitter angle, SWAP, and SWAP\texorpdfstring{$^2$}{2}}
\label{sec:swap-glossary}

Three related quantities recur; we fix their definitions here.

\emph{Beam-splitter angle.} $\theta \equiv \int g(t)\,dt$ is the pulse area of the beam-splitter drive envelope $g(t)$, setting the generator angle of $\hat H_{\rm BS} = g(\hat b^\dagger \hat a_t + \hat b\, \hat a_t^\dagger) = 2g\,\hat J_x$ under the Jordan--Schwinger map (Sec.~\ref{sec:sm-su2}).

\emph{Elementary SWAP.}  One ``SWAP'' is a single beam-splitter pulse with $\theta = \pi/2$ (pulse area $\pi/2$), duration $T_{\rm SWAP} = \pi/(2\bar g)$ for effective coupling $\bar g$.  On single-mode inputs it maps $\hat a_t \to -i\hat b$ and $\hat b \to -i\hat a_t$, exchanging the two-mode amplitudes with a per-photon $-i$ phase.  It is a physical exchange of photon populations, \emph{not} a logical SWAP gate.

\emph{SWAP\texorpdfstring{$^2$}{2}.}  Two consecutive SWAPs (total pulse area $\pi$, total beam-splitter angle $\theta = \pi$, total duration $2\,T_{\rm SWAP}$).  Under the SU(2) argument of Sec.~\ref{sec:sm-su2}, SWAP$^2\ket{m,n}_{b,t}$ picks up $(-1)^{m+n}$~\cite{YurkeMcCallKlauder1986,CamposSalehTeich1989}; on $\ket{0,n}$ this becomes $(-1)^n$, and on a coherent state $\ket{\alpha}$ it realises $\ket{\alpha}\to\ket{-\alpha}$ --- i.e., the cat $X$ gate.  In particular, SWAP$^2 \neq \openone$.

\emph{cSWAP\texorpdfstring{$^2$}{2}.}  A SWAP$^2$ made conditional on the control-cat state via cross-Kerr.  When the control is in $\ket{0}$ (ON) the SWAP$^2$ executes; when the control is in $\ket{2\alpha_c}$ (OFF) the cross-Kerr detunes the bin and no exchange occurs.  In the displaced-basis cat logical space this implements a controlled-$X$.

\emph{$\chi$-echo halves.}  The $\chi$-echo (Sec.~\ref{sec:sm-chiecho}) splits SWAP$^2$ into two elementary SWAPs of duration $T_{\rm SWAP}$ each with the sign of $\chi$ reversed between them.  ``Per $\chi$-echo half'' therefore means one SWAP.

\emph{Full CNOT duration.}  The full protocol wraps SWAP$^2$ in a control-cat displacement echo, giving $T_{\rm CX} = 4\,T_{\rm SWAP}$ (two SWAPs plus a matched idle window; Sec.~\ref{sec:sm-dispecho}).

\subsection{Notation summary}
\label{sec:notation-summary}

\noindent
\begin{tabular}{@{}l@{\ }p{0.66\columnwidth}@{}}
$p_X^{(c)}$, $p_X^{(t)}$ & Control / target bit-flip error \\
$p_Z^{(c)}$, $p_Z^{(t)}$ & Control / target phase-flip error \\
$p_L$ & Logical error per syndrome cycle \\
$\bar{n} = |\alpha|^2$ & Cat photon number (data qubits) \\
$\bar{n}_a$ & Ancilla cat photon number \\
$\bar{n}_{\rm th}$ & Thermal population of bin mode \\
$\theta$ & Beam-splitter angle $\equiv \int g(t)\,dt$ (pulse area) \\
SWAP & Elementary pulse: $\theta = \pi/2$, duration $T_{\rm SWAP} = \pi/(2\bar g)$ \\
SWAP$^2$ & Two SWAPs, total $\theta = \pi$, $(-1)^{\hat n}$ phase; realises $X$ on cat \\
$T_{\rm CX}$ & Full CNOT duration $= 4\,T_{\rm SWAP}$ \\
$|\mathcal{C}_\pm\rangle$ & Even/odd parity cat states \\
$\hat{J}_{++}$, $\hat{J}_{+-}$ & Stabilizer conserved quantities for parity and coherence \\
\end{tabular}

\section{Beam-splitter general solution (resonant and detuned)}
\label{sec:sm-beamsplitter}
We collect the general solutions of the two-mode beam-splitter
Hamiltonian used throughout this work.  Consider the time-dependent
Hamiltonian in the rotating frame of both target and bin mode:
\begin{equation}\label{eq:Hbs}
\hat{H}(t) = \Delta\,\hat{b}^\dagger\hat{b}
  + g(t)\bigl(\hat{b}^\dagger\hat{a}_t + \hat{b}\,\hat{a}_t^\dagger\bigr),
\end{equation}
where $\Delta$ is the detuning of the bin mode from the beam-splitter
drive and $g(t)$ is a real, time-dependent coupling envelope.
Equation~\eqref{eq:Hbs} conserves the total photon number
$\hat{N} = \hat{a}_t^\dagger\hat{a}_t + \hat{b}^\dagger\hat{b}$
and therefore acts independently on each fixed-$N$ subspace.

\subsection{Resonant case ($\Delta = 0$)}

When the drive is on resonance the Hamiltonian reduces to
$\hat{H}_{\mathrm{ON}}(t) = g(t)(\hat{b}^\dagger\hat{a}_t + \hat{b}\,\hat{a}_t^\dagger)$.
The Heisenberg equations of motion are
\begin{equation}
i\dot{\hat{a}}_t = g(t)\,\hat{b},\qquad
i\dot{\hat{b}} = g(t)\,\hat{a}_t,
\end{equation}
which decouple into the normal modes
$\hat{c}_\pm = (\hat{a}_t \pm \hat{b})/\sqrt{2}$ satisfying
$i\dot{\hat{c}}_\pm = \pm g(t)\,\hat{c}_\pm$.
The solution is
\begin{align}
\hat{a}_t(T) &= \cos\theta(T)\;\hat{a}_t(0) - i\sin\theta(T)\;\hat{b}(0),\notag\\
\hat{b}(T)   &= -i\sin\theta(T)\;\hat{a}_t(0) + \cos\theta(T)\;\hat{b}(0),
\label{eq:BS_resonant}
\end{align}
where the accumulated beam-splitter angle is
\begin{equation}\label{eq:theta_def}
\theta(T) = \int_0^T g(t')\,dt'.
\end{equation}
One elementary SWAP corresponds to $\theta = \pi/2$ (pulse area $\pi/2$, duration $T_{\rm SWAP} = \pi/(2\bar g)$; Sec.~\ref{sec:swap-glossary}), yielding $\hat{a}_t \to -i\hat{b}$, $\hat{b}\to -i\hat{a}_t$.
Two such SWAPs give a full SWAP$^2$: $\hat{a}_t\to -\hat{a}_t$,
$\hat{b}\to -\hat{b}$.  Applied to a coherent-state input
$|0,\alpha\rangle_{b,t}$, the first SWAP gives
$|{-}i\alpha,0\rangle_{b,t}$ and the full SWAP$^2$ cycle gives
$|0,-\alpha\rangle_{b,t}$, realizing $|\alpha\rangle \to |{-}\alpha\rangle$
on the target --- i.e., a bit flip ($X$ gate) on the cat qubit.

\subsection{Detuned case ($\Delta \neq 0$)}
\label{sec:sm-beamsplitter-detuned}

For constant detuning and coupling, $\hat{H} = \Delta\,\hat{b}^\dagger\hat{b}
+ g(\hat{b}^\dagger\hat{a}_t + \hat{b}\,\hat{a}_t^\dagger)$, the
Heisenberg equations are
\begin{equation}
i\dot{\hat{a}}_t = g\,\hat{b},\qquad
i\dot{\hat{b}} = \Delta\,\hat{b} + g\,\hat{a}_t.
\end{equation}
Moving to the interaction picture with respect to $\Delta\,\hat{b}^\dagger\hat{b}$
and defining $\tilde{b} = \hat{b}\,e^{i\Delta t}$, one obtains
$i\dot{\tilde{b}} = g\,e^{i\Delta t}\hat{a}_t$ and
$i\dot{\hat{a}}_t = g\,e^{-i\Delta t}\tilde{b}$, which combine into
a second-order equation for $\hat{a}_t$:
\begin{equation}\label{eq:at_ode}
\ddot{\hat{a}}_t + i\Delta\,\dot{\hat{a}}_t + g^2\,\hat{a}_t = 0.
\end{equation}
The characteristic frequencies are
$\lambda_\pm = (-\Delta \pm \Omega_R)/2$, where the generalized Rabi
frequency is $\Omega_R = \sqrt{\Delta^2 + 4g^2}$.  With initial
conditions $\hat{a}_t(0)$ and $\hat{b}(0) = (i\dot{\hat{a}}_t(0))/g$,
the solution is
\begin{align}
\hat{a}_t(t) &= e^{-i\Delta t/2}\Bigl[
  \cos\tfrac{\Omega_R t}{2}\;\hat{a}_t(0)
  + \frac{i\Delta}{\Omega_R}\sin\tfrac{\Omega_R t}{2}\;\hat{a}_t(0)
  \notag\\
  &\quad - \frac{2ig}{\Omega_R}\sin\tfrac{\Omega_R t}{2}\;\hat{b}(0)
\Bigr],\label{eq:at_detuned}\\[4pt]
\hat{b}(t) &= e^{-i\Delta t/2}\Bigl[
  - \frac{2ig}{\Omega_R}\sin\tfrac{\Omega_R t}{2}\;\hat{a}_t(0)
  \notag\\
  &\quad + \cos\tfrac{\Omega_R t}{2}\;\hat{b}(0)
  - \frac{i\Delta}{\Omega_R}\sin\tfrac{\Omega_R t}{2}\;\hat{b}(0)
\Bigr].\label{eq:b_detuned}
\end{align}
In the far-detuned limit $|\Delta| \gg g$, the Rabi frequency
$\Omega_R \approx |\Delta| + 2g^2/|\Delta|$, and the transfer
probability $P_{t\to b} = (2g/\Omega_R)^2\sin^2(\Omega_R t/2) \approx
(2g/\Delta)^2\sin^2(\Omega_R t/2)$ is strongly suppressed.  The
dominant effect on the target is an AC Stark shift
$\delta\omega_t = -g^2/\Delta$ from virtual exchange with the bin,
which accumulates a deterministic phase
\begin{equation}\label{eq:stark_phase}
\varphi_{\rm Stark} = \frac{g^2 T}{\Delta}
\end{equation}
on the target cat in the OFF branch.  In this work, the $\chi$-echo
protocol (Sec.~\ref{sec:sm-chiecho}) cancels this Stark phase to
leading order by flipping the sign of $\Delta$ at the gate midpoint.
Alternatively, without a $\chi$-echo, the Stark phase can be corrected
by applying a small detuning to the second SWAP half-pulse in the ON
branch, so that the ON-branch target acquires the same phase-space
rotation as the OFF branch; the cat qubit logical axis is then
redefined to absorb the common rotation, requiring no additional
correction pulse.

\subsection{Time-dependent coupling}

For a shaped pulse $g(t)$ with detuning $\Delta$ (as used in the gate),
the mode amplitudes obey
\begin{equation}\label{eq:coupled_ode}
i\frac{d}{dt}\begin{pmatrix}\hat{a}_t\\\hat{b}\end{pmatrix}
= \begin{pmatrix} 0 & g(t) \\ g(t) & \Delta \end{pmatrix}
\begin{pmatrix}\hat{a}_t\\\hat{b}\end{pmatrix}.
\end{equation}
When the adiabatic condition $|\dot{g}| \ll \Omega_R^2$ is satisfied,
the instantaneous eigenstates of the coupling matrix are
$|\pm\rangle \propto (2g,\,\Delta \pm \Omega_R)^T$ with eigenvalues
$\lambda_\pm = (\Delta \pm \Omega_R)/2$.  The adiabatic solution gives
the transfer probability at time $T$ as
$P_{t\to b}(T) \approx (2g_{\max}/\Omega_{R,\max})^2
\sin^2[\frac{1}{2}\int_0^T\!\Omega_R(t')\,dt']$, recovering the
resonant result for $\Delta = 0$ and the far-detuned suppression for
$\Delta \gg g_{\max}$.

In the gate, the ON branch ($\Delta = 0$) uses a shaped pulse with
area~\eqref{eq:theta_def} calibrated to $\theta = \pi/2$ per echo
half, while the OFF branch (average detuning
$\bar{\Delta} = \chi\langle\hat{n}_c\rangle = 4\chi|\alpha_c|^2$)
relies on the far-detuned suppression to keep the bin in vacuum.
The peak transient bin excitation at the mean detuning is
\begin{equation}\label{eq:nbar_res}
\bar{n}_b^{\rm peak} \approx
\biggl(\frac{2g_{\max}}{4\chi|\alpha_c|^2}\biggr)^2,
\end{equation}
which sets the scale of the peak off-resonant excitation during the
pulse.  The $\chi$-echo (Sec.~\ref{sec:sm-chiecho}) refocuses the
cross-Kerr phase from this transient population, and for a
Gaussian-ramped pulse the residual bin population after the gate is
exponentially suppressed below this value
(Sec.~\ref{sec:sm-drag}).

\section{SU(2) equivalence, geometric phase, and \texorpdfstring{CR$_X$}{CR\_X} analogy}
\label{sec:sm-su2}
\subsection{Fock-state phase under SWAP and SWAP$^2$}

The origin of the $X$ gate from SWAP$^2$ can be seen directly from the
Fock-state phases acquired during a beam-splitter exchange.
We derive the phase acquired by an arbitrary two-mode Fock state
under the unitary for one elementary SWAP, $U = e^{-i\pi\hat{J}_x}$.  The
Heisenberg relations (Sec.~\ref{sec:sm-beamsplitter},
Eq.~\ref{eq:BS_resonant}) at $\theta = \pi/2$ give
$U^\dagger \hat{a}_t\, U = -i\hat{b}$ and
$U^\dagger \hat{b}\, U = -i\hat{a}_t$.  Taking the adjoint:
\begin{equation}\label{eq:Udag_conj}
U\hat{a}_t^\dagger U^\dagger = -i\hat{b}^\dagger,\qquad
U\hat{b}^\dagger U^\dagger = -i\hat{a}_t^\dagger.
\end{equation}
Since $|m,n\rangle_{b,t} = (\hat{b}^\dagger)^m (\hat{a}_t^\dagger)^n
|0,0\rangle / \sqrt{m!\,n!}$, we apply $U$ by conjugating each
creation operator:
\begin{align}
U|m,n\rangle_{b,t}
&= \frac{(-i\hat{a}_t^\dagger)^m\;(-i\hat{b}^\dagger)^n}
        {\sqrt{m!\,n!}}\;|0,0\rangle,
\end{align}
giving
\begin{equation}\label{eq:fock_swap}
\boxed{\;U|m,n\rangle_{b,t} = (-i)^{m+n}\,|n,m\rangle_{b,t}.\;}
\end{equation}
Each photon contributes one factor of $(-i)$ while the mode
populations exchange.

For an initial state with the bin in vacuum ($m = 0$), a single SWAP
gives $U|0,n\rangle = (-i)^n|n,0\rangle$, and a second application:
\begin{equation}\label{eq:swap2_fock}
U^2|0,n\rangle_{b,t} = (-1)^n\,|0,n\rangle_{b,t}.
\end{equation}
This is the $(-1)^n$ geometric phase responsible for the
$|\alpha\rangle \to |{-}\alpha\rangle$ mapping.  Applied to a coherent
state by linearity:
\begin{align}
|\alpha\rangle = e^{-|\alpha|^2/2}\sum_n
\frac{\alpha^n}{\sqrt{n!}}\,|n\rangle
&\xrightarrow{\;\text{SWAP}^2\;}
|{-}\alpha\rangle.
\label{eq:alpha_to_minus_alpha}
\end{align}
For the cat qubit logical states
$|\mathcal{C}_\pm\rangle \propto |\alpha\rangle \pm |{-}\alpha\rangle$
(even/odd Fock components):
\begin{equation}
|\mathcal{C}_+\rangle \to |\mathcal{C}_+\rangle, \qquad
|\mathcal{C}_-\rangle \to -|\mathcal{C}_-\rangle,
\end{equation}
confirming SWAP$^2$ acts as $\hat{X}$ on the cat qubit
(a bit flip in the $|\pm\alpha\rangle$ computational basis).

\subsection{Jordan-Schwinger map and SU(2) structure}

An alternative geometric understanding follows from the SU(2) structure
of the beam-splitter.  The Hamiltonian
$\hat{H}_{\rm BS} = g(\hat{b}^\dagger\hat{a}_t + \hat{b}\,\hat{a}_t^\dagger)$
conserves the total photon number
$\hat{N} = \hat{a}_t^\dagger\hat{a}_t + \hat{b}^\dagger\hat{b}$.
Within the fixed-$N$ subspace, the two-mode system admits an SU(2)
structure via the Jordan-Schwinger map~\cite{Schwinger1965}:
\begin{align}
\hat{J}_x &= \tfrac{1}{2}(\hat{b}^\dagger\hat{a}_t
+ \hat{b}\,\hat{a}_t^\dagger), \label{eq:Jx}\\
\hat{J}_y &= \tfrac{1}{2i}(\hat{b}^\dagger\hat{a}_t
- \hat{b}\,\hat{a}_t^\dagger), \label{eq:Jy}\\
\hat{J}_z &= \tfrac{1}{2}(\hat{a}_t^\dagger\hat{a}_t
- \hat{b}^\dagger\hat{b}), \label{eq:Jz}
\end{align}
satisfying $[\hat{J}_i, \hat{J}_j] = i\epsilon_{ijk}\hat{J}_k$.
The total spin quantum number is $j = N/2$, and the beam-splitter
Hamiltonian is $\hat{H}_{\rm BS} = 2g\,\hat{J}_x$, generating
rotations about the $x$-axis:
\begin{equation}\label{eq:U_Jx}
U(\theta) = e^{-i\theta\,\hat{J}_x}, \qquad \theta = 2gt.
\end{equation}
In a spin-$j = N/2$ representation, a $\pi$ rotation about
$\hat{J}_x$ (corresponding to one elementary SWAP, $\theta = \pi$) maps
$|j,m\rangle \to (-1)^{j-m}|j,-m\rangle$, which for $j = N/2 = n/2$
(bin starting in vacuum) gives the same $(-1)^n$ phase derived above.

\subsection{\texorpdfstring{CR$_X$}{CR\_X} analogy: spin-$1/2$ vs.\ spin-$N/2$}

The cat-transmon controlled-rotation-X (CR$_X$) gate~\cite{Hann2024} conditionally
rotates a transmon (spin-$1/2$) via a dispersive shift from a storage
cat (the analog of our control cat).  The Jordan-Schwinger map of the
previous subsection allows us to draw an exact analogy: The VCB gate
conditionally rotates the bin-target spin-$N/2$ system via the
cross-Kerr shift from the control cat.
The correspondence can be made algebraically exact.
For the CR$_X$ gate, the dispersive interaction
$\hat{H}_\chi = \chi_e\,|e\rangle\langle e|\,\hat{n}_c$ gives, using
$|e\rangle\langle e| = \hat{J}_z^{(1/2)} + 1/2$:
\begin{align}\label{eq:H_CRX}
\hat{H}_{\mathrm{CR}_X}
&= \chi_e\,\hat{n}_c\,\hat{J}_z^{(1/2)}
  + \Omega\,\hat{J}_x^{(1/2)}
  + \underbrace{\tfrac{\chi_e}{2}\hat{n}_c}_{\text{ground-state shift}}.
\end{align}
For the VCB, using $\hat{n}_b = N/2 - \hat{J}_z$:
\begin{equation}\label{eq:H_VCB_spin}
\hat{H}_{\rm VCB}
= -\chi\,\hat{n}_c\,\hat{J}_z + 2g\,\hat{J}_x
  + \underbrace{\tfrac{\chi}{2}\hat{n}_c N}_{\text{vacuum shift}}.
\end{equation}
The vacuum shift is the cross-Kerr energy when the bin is empty
(all photons in target); within each fixed-$N$ sector it is a
deterministic control frequency shift absorbed into a rotating frame:
\begin{equation}\label{eq:H_VCB_eff}
\hat{H}_{\rm VCB}^{\rm eff}
= -\chi\,\hat{n}_c\,\hat{J}_z + 2g\,\hat{J}_x.
\end{equation}

The two effective Hamiltonians have identical structure under
$\hat{J}^{(1/2)} \to \hat{J}^{(N/2)}$, $\Omega \to 2g$,
$\chi_e \to -\chi$.  In both, the control photon number sets the
detuning of a driven spin rotation: $n_c = 0$ gives free rotation,
$n_c \gg 1$ suppresses it.

Table~\ref{tab:crx_comparison} summarizes the correspondence.

\begin{table}[h]
\caption{\label{tab:crx_comparison}
Comparison of CR$_X$ and VCB gate mechanisms.}
\vspace{16pt}
\begin{tabular}{l l l}
\hline\hline
Property & CR$_X$~\cite{Hann2024} & VCB (this work) \\
\hline
Target mode & Transmon & \{target cat, bin\} \\
Spin representation & Spin-$1/2$ & Spin-$N/2$ \\
Rotation generator & $\hat{\sigma}_x$ & $\hat{J}_x$ \\
Gate rotation & $\pi$ (spin flip) & SWAP$^2$ ($\pi$ BS angle) \\
Gate action & $|g\rangle \to |e\rangle$ & $|\alpha\rangle \to |{-}\alpha\rangle$ \\
Nonlinear coupling & Dispersive shift $\chi_e$ & Cross-Kerr $\chi$ \\
\hline\hline
\end{tabular}
\end{table}

\subsection{VCB as a superposition of \texorpdfstring{CR$_X$}{CR\_X} gates}

The effective Hamiltonian~\eqref{eq:H_VCB_eff} is block-diagonal in
the control Fock basis: For fixed $n_c$, the target-bin system evolves
under
\begin{equation}\label{eq:H_nc}
\hat{H}^{(n_c)} = -\chi\,n_c\,\hat{J}_z + 2g(t)\,\hat{J}_x,
\end{equation}
structurally identical to the CR$_X$ Hamiltonian with $n_c$ playing the
role of the CR$_X$ storage photon number.  The control cat creates a
superposition of such Hamiltonians at different detunings.

Since the beam-splitter conserves $N = n_t + n_b$, the target-bin
Hilbert space decomposes as
\begin{equation}\label{eq:hilbert_decomp}
\mathcal{H}_{tb} = \bigoplus_{N=0}^{\infty}\, \mathcal{H}_N,
\qquad \dim\mathcal{H}_N = N+1,
\end{equation}
where $\mathcal{H}_N$ carries spin-$j = N/2$.  The VCB unitary
performs an independent CR$_X$ gate in each subspace:
\begin{equation}\label{eq:superposition_CRX}
U_{\rm VCB} = \bigoplus_{N}\, U_{\mathrm{CR}_X}^{(j = N/2)}.
\end{equation}
The target cat populates multiple $N$ sectors (with bin in vacuum,
$N = n_t$), so the gate acts as a coherent superposition of CR$_X$ gates
at different spin representations.

\section{Target cat bias preservation during SWAP}
\label{sec:sm-biasswap}
\label{sec:sm-bitflip}%

In this section we prove that the SWAP$^2$ gate preserves the exponential bit-flip suppression of the cat qubit under realistic imperfections: single-photon loss, self-Kerr nonlinearity, and thermal population.

\subsection{Under single-photon loss ($T_1$)}
\label{sec:sm-swap-T1}

We prove that single-photon loss at any point during the beam-splitter
SWAP cycle preserves the bare cat-qubit bit-flip suppression
$p_X \approx \bar{n}\,\kappa_1 T\,e^{-2\bar{n}}$.

Consider the resonant beam-splitter Hamiltonian
$\hat{H}_{\rm BS} = g(\hat{b}^\dagger\hat{a}_t
+ \hat{b}\,\hat{a}_t^\dagger)$, which conserves the total photon number
$\hat{N} = \hat{a}_t^\dagger\hat{a}_t + \hat{b}^\dagger\hat{b}$.
The unitary $U(\theta) = e^{-i\theta(\hat{b}^\dagger\hat{a}_t
+ \hat{b}\,\hat{a}_t^\dagger)}$ maps the initial state
$|0,\alpha\rangle_{b,t}$ to
\begin{equation}\label{eq:state_theta}
|\psi(\theta)\rangle = U(\theta)\,|0,\alpha\rangle_{b,t}
= |{-}i\alpha\sin\theta,\;\alpha\cos\theta\rangle_{b,t},
\end{equation}
using the fact that the beam-splitter maps coherent states to
coherent states.

A single-photon loss event in the target mode at angle~$\theta$ gives
$\hat{a}_t|\psi(\theta)\rangle = \alpha\cos\theta\,|\psi(\theta)\rangle$.
The bit-flip matrix element, defined as the overlap of the post-loss state with
the orthogonal logical state propagated to the same angle, is
\begin{align}
\mathcal{M}_t(\theta)
&= \langle\bar\psi(\theta)|\,\hat{a}_t\,|\psi(\theta)\rangle
\notag\\
&= \alpha\cos\theta\;
   \langle-\alpha\cos\theta|\alpha\cos\theta\rangle_t\;
   \langle i\alpha\sin\theta|{-}i\alpha\sin\theta\rangle_b
\notag\\
&= \alpha\cos\theta\;e^{-2\alpha^2\cos^2\theta}\,e^{-2\alpha^2\sin^2\theta},
\end{align}
which simplifies to
\begin{equation}\label{eq:Mt_result}
\mathcal{M}_t(\theta) = \alpha\cos\theta\;e^{-2\alpha^2}.
\end{equation}
The same calculation for bin loss gives
\begin{equation}\label{eq:Mb_result}
\mathcal{M}_b(\theta) = -i\alpha\sin\theta\;e^{-2\alpha^2}.
\end{equation}
The total two-mode bit-flip susceptibility is therefore
\begin{equation}\label{eq:sum_rule}
|\mathcal{M}_t(\theta)|^2 + |\mathcal{M}_b(\theta)|^2
= \alpha^2 e^{-4\alpha^2},
\end{equation}
independent of~$\theta$.  For comparison, the idle (by ``idle'' we mean free evolution without two-photon stabilization, followed by projection onto the cat basis to extract errors; see Sec.~\ref{sec:sm-error-defs}) cat-qubit matrix
element is
$\mathcal{M}_{\rm idle} = \langle{-}\alpha|\hat{a}|\alpha\rangle
= \alpha\,e^{-2\alpha^2}$, giving
$|\mathcal{M}_{\rm idle}|^2 = \alpha^2 e^{-4\alpha^2}$---identical to
Eq.~\eqref{eq:sum_rule}.  Thus single-photon loss during the SWAP is
no worse than during idle at every point in the cycle.

We verify this numerically by simulating the full master equation for
the two-mode (target + bin) SWAP$^4$ gate ($X^2 = I$) and
comparing against a single-mode target cat under idling and continuous
two-photon stabilization (TPS), all with identical loss and dephasing
rates.
Figure~\ref{fig:swap_bias_T1}(a) confirms that all three protocols
give the same exponential bit-flip suppression; the grey line shows
the analytic rate $\bar{n}\,\kappa_1 T\,e^{-2\bar{n}}$.
Figure~\ref{fig:swap_bias_T1}(b) shows the phase-flip grows linearly
as $\bar{n}\,\kappa_1 T$, with the SWAP$^4$ value ${\sim}20\%$ larger
than idle from additional bin-mode photon loss.

\begin{figure}[t]
\includegraphics[width=\columnwidth]{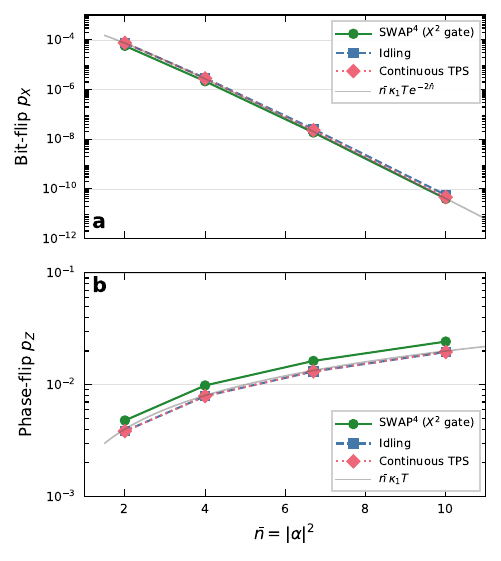}
\caption{\label{fig:swap_bias_T1}
Bias preservation of the SWAP$^4$ cycle under single-photon loss
($T_1 = 500$\,$\mu$s, $\kappa_\phi = \kappa_1$, gate time $T = 1$\,$\mu$s).
(a)~Bit-flip probability~$p_X$: SWAP$^4$ (green), idle (blue), and
TPS (red) all exhibit the same exponential suppression.
Grey line: $\bar{n}\,\kappa_1 T\,e^{-2\bar{n}}$.
(b)~Phase-flip probability~$p_Z$.  Grey line: $\bar{n}\,\kappa_1 T$.
}
\end{figure}

\subsection{Under Kerr-matching ($K_b = K_t$)}
\label{sec:sm-selfkerr-swap}

The analytic proof above assumes a purely linear beam-splitter.
In practice, both modes possess self-Kerr nonlinearities.  Under the
Kerr-matching condition $K_b = K_t \equiv K$, the SWAP transfers the
coherent state between symmetric modes, so the nonlinear phase
accumulation is balanced.
We verify numerically that Kerr-matching does not degrade the
bit-flip suppression by comparing a cat qubit idling under self-Kerr
(i.e., evolving under $(K/2)\,\hat{a}^{\dagger 2}\hat{a}^2$ without
stabilization, then projecting onto the cat basis to extract errors)
with one undergoing the full SWAP$^2$ cycle.

Figure~\ref{fig:swap_bias_kerr}(a) shows $p_X$ versus $\bar{n}$ for
three values of~$Kt$.  At all values, the SWAP$^2$ bit-flip is
\emph{lower} than idle, because the SWAP distributes the photon
population across two modes, reducing the effective $n^2$ contribution
of the self-Kerr at any instant.

Figure~\ref{fig:swap_bias_kerr}(b) shows the phase-flip error.
Unlike idle (where self-Kerr preserves parity, giving $p_Z = 0$),
SWAP$^2$ acquires a small phase-flip from incomplete Fock-state
transfer under Kerr, but this remains below $2\times 10^{-5}$ for
$Kt \leq 0.06$.  At our operating point
($\chi/K \gtrsim 1000$, $T \sim 50$--$100$\,ns, $Kt \lesssim 0.03$),
this is negligible compared to photon-loss dephasing.

\begin{figure}[t]
\includegraphics[width=\columnwidth]{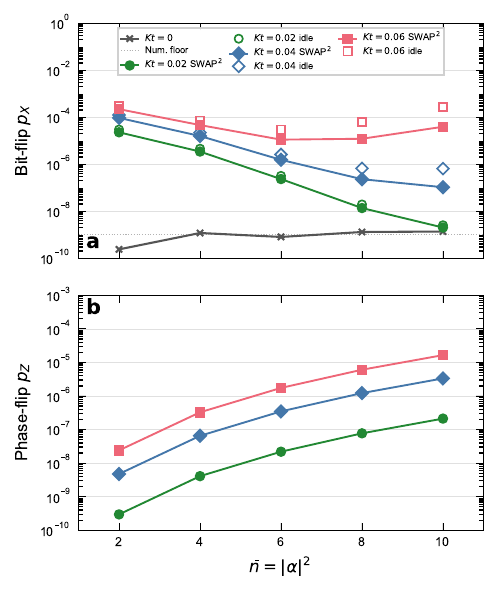}
\caption{\label{fig:swap_bias_kerr}
Target bit-flip and phase-flip under Kerr-matching
($K_b = K_t$).
(a)~Bit-flip probability~$p_X$ versus $\bar{n}$.
Open markers: idle; filled markers: SWAP$^2$.
The SWAP$^2$ bit-flip is lower than idle at all~$Kt$.
(b)~Phase-flip probability~$p_Z$ from the SWAP$^2$ cycle.
Remains below $2\times 10^{-5}$ for $Kt \leq 0.06$.
}
\end{figure}

\subsection{Under thermal control photons}
\label{sec:sm-thermal-target-on}
In this section we analyze the target ON bit-flip error arising from residual thermal population in the control mode.  Errors are extracted by projection onto the cat basis as defined in Sec.~\ref{sec:sm-error-defs}.

When the control cat has the assumed residual thermal population
$\bar{n}_{\rm th} \approx 10^{-4}$, the displaced control state is
not a pure vacuum but a thermal mixture
$\hat{\rho}_c = \sum_{n=0}^{\infty} p_n |n\rangle\langle n|$ with
$p_n = \bar{n}_{\rm th}^n/(1+\bar{n}_{\rm th})^{n+1}$.
In the limit $\bar{n}_{\rm th} \ll 1$ (which is well satisfied at
$10^{-4}$), this is dominated by the $n_c = 0$ and $n_c = 1$
components: $p_0 \approx 1 - \bar{n}_{\rm th}$,
$p_1 \approx \bar{n}_{\rm th}$, with $p_{n \geq 2} = O(\bar{n}_{\rm th}^2)$
negligible.  We therefore approximate the thermal state as
$\hat{\rho}_c \approx (1-\bar{n}_{\rm th})|0\rangle\langle 0|
+ \bar{n}_{\rm th}\,|1\rangle\langle 1|$ and analyze the
$n_c = 1$ component separately.

The $n_c = 1$ component activates the cross-Kerr detuning
$\Delta = \chi$ on the bin, shifting the beam-splitter interaction
off resonance.  The SWAP$^2$ cycle then no longer completes a full
photon exchange between target and bin, and the target coherent
state does not return perfectly to its starting amplitude---producing
a bit-flip error.

The $\chi$-echo partially refocuses this detuning, but only in the
limit $g \gg \chi$ where the echo halves are nearly identical.
When $g \sim \chi$, the detuned SWAP dynamics are no longer
time-symmetric and the echo is less effective.

\textit{Simulation setup.}---We isolate this channel by simulating
the two-mode subsystem $\{\hat{b},\,\hat{a}_t\}$ under
\begin{equation}\label{eq:H_detuned_swap}
\hat{H}_{\rm det} = \Delta\,\hat{b}^\dagger\hat{b}
+ g(t)\bigl(\hat{a}_t^\dagger\hat{b} + \hat{a}_t\,\hat{b}^\dagger\bigr),
\end{equation}
with $\Delta = \chi$ (the shift induced by $n_c = 1$) and the
$\chi$-echo implemented by flipping the sign of $\Delta$ at the pulse
midpoint.  The Gaussian pulse shape and normalization
$\int_0^T g(t)\,dt = \pi/2$ per elementary SWAP are as in
Sec.~\ref{sec:sm-drag}; convergence details are in
Sec.~\ref{sec:sm-simulations}.  The initial state has the bin in
vacuum and the target in $|\alpha_t\rangle$; after the two-pulse
sequence we project onto the asymptotic cat basis
(Sec.~\ref{sec:sm-error-defs}) and report the bit-flip
$p_X = |\rho_{1,1}|$, where $\rho_{1,1}$ is the population of the
wrong-cat pole $|{-}\alpha_t\rangle$ after the intended
$|\alpha_t\rangle \to |{-}\alpha_t\rangle$ flip.
Parameters: $\chi/2\pi = 4$~MHz, $\sigma = 0.18$; we sweep
$\bar{n}_t = |\alpha_t|^2 \in \{2,4,6,8,10\}$ at four
gate times $T \in \{25, 50, 75, 100\}$~ns per elementary SWAP.

\textit{Results.}---Figure~\ref{fig:thermal_target_on}(b) shows the
per-thermal-photon target bit-flip versus $\bar{n}_t$.  Shorter gates
have larger $g_{\max}/\chi$, making the detuning perturbative
[panel (a)]: at $T = 25$~ns ($g_{\max}/\chi \approx 5.7$) the error
falls below $10^{-6}$ for $\bar{n}_t \geq 4$.  At the
syndrome-extraction point $T = 50$~ns ($g_{\max}/\chi \approx 2.9$)
the per-thermal-photon error is ${\sim}3\times 10^{-5}$ at
$\bar{n}_t = 4$, giving a net contribution
${\sim}\bar{n}_{\rm th}\cdot 3\times 10^{-5} \approx 3\times 10^{-9}$.
At the transversal CNOT point $T = 100$~ns
($g_{\max}/\chi \approx 1.4$) the per-photon error rises to
${\sim}10^{-4}$ at $\bar{n}_t = 8$ (net ${\sim}10^{-8}$).
In all cases this channel is negligible.

The gate time therefore trades this channel (favoring large
$g_{\max}/\chi$, short $T$) against OFF-branch control bit-flip
(favoring small $g_{\max}/\bar{\Delta}_{\rm OFF}$, long $T$;
Sec.~\ref{sec:sm-drag}).  At $\chi/2\pi = 4$~MHz, $T = 50$~ns balances
both.

\begin{figure}[t]
\centering
\includegraphics[width=\columnwidth]{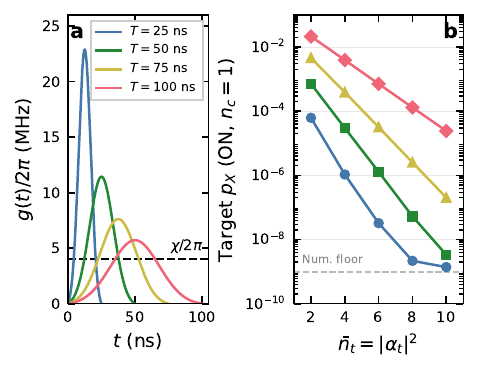}
\caption{\label{fig:thermal_target_on}
Target ON-branch bit-flip from thermal control photons.
(a)~Beam-splitter pulse envelopes $g(t)/2\pi$ for four elementary
SWAP durations $T$.  Black dashed line: cross-Kerr detuning
$\chi/2\pi = 4$~MHz seen by the $n_c = 1$ component.  Shorter
pulses have $g_{\max} \gg \chi$, making the detuning perturbative.
(b)~Target bit-flip error conditioned on $n_c = 1$ versus
$\bar{n} = |\alpha|^2$.  Colors match panel~(a).
The error decreases with both $\bar{n}_t$ (the exponential bit-flip
suppression $\sim e^{-2|\alpha_t|^2}$ makes the cat more robust to
small coherent-state distortions at larger size) and shorter $T$
(larger $g/\chi$).
The net contribution to the target bit-flip is this value multiplied
by $\bar{n}_{\rm th} \approx 10^{-4}$.
Grey dashed: numerical floor ($10^{-9}$).
Simulation: two-mode Schr\"odinger evolution under
Eq.~\eqref{eq:H_detuned_swap} with $\chi$-echo
($\Delta \to -\Delta$ at midpoint), $\sigma = 0.18$,
$\chi/2\pi = 4$~MHz.
}
\end{figure}

\section{Control cat bias preservation with cross-Kerr coupling to bin}
\label{sec:sm-control}
In this section we analyze the control cat's bit-flip error arising from its cross-Kerr coupling to the bin mode during the gate.  Errors are extracted by projection onto the cat basis (Sec.~\ref{sec:sm-error-defs}).

This section is organized by branch and error mechanism.

\subsection{Control ON branch}

In the ON branch, the control is displaced to $|0\rangle$ (vacuum).
Since the cross-Kerr Hamiltonian $\chi\,\hat{n}_c\hat{n}_b$ vanishes
at $\hat{n}_c = 0$, the bin dynamics are completely decoupled from the
control and cannot induce a control bit-flip regardless of the bin
state.  Furthermore, $|0\rangle$ is an eigenstate of the photon loss
operator ($\hat{a}|0\rangle = 0$), so the control ON branch is
completely inert during the gate.

\subsection{Control OFF branch: resilience to bin $T_1$ loss}
\label{sec:T1_comparison}

In the OFF branch, the bin is populated only off-resonantly by the
beam-splitter drive detuned by $\bar{\Delta} = 4\chi|\alpha_c|^2$.
The bin can be thought of as being continuously driven: Any photon
lost from the bin is quickly replenished by the drive, with negligible
change in the bin's steady-state amplitude.  Since the drive replenishes lost photons, the bin remains in
essentially the same coherent state regardless of $T_1$, and the
control OFF state cross-Kerr couples to an unchanged bin field---so
the control bit-flip remains exponentially suppressed.

We verify this numerically by simulating the reduced control--bin
system (with the target mode replaced by a classical amplitude
$\alpha_t$) under the $\chi$-echo protocol with Lindblad dissipation
at rates corresponding to bin $T_1 = 10$, $100$, and $1000$~$\mu$s.
The solid lines in Fig.~\ref{fig:control_bin_T1} show the control OFF
bit-flip error as a function of $|\alpha|^2$ for each $T_1$ value.
The exponential suppression is maintained across all $T_1$ values:
Even at $T_1 = 10$~$\mu$s ($\kappa\,T_{\rm gate} \sim 2\times 10^{-2}$),
the bit-flip error follows the same $e^{-c|\alpha|^2}$ scaling as the
unitary case, with no $\bar{n}$-independent floor.

\subsection{Control OFF branch: thermal excitation sets a bit-flip floor}
\label{sec:thermal_sensitivity}

Unlike the off-resonant bin population (which is drive-protected),
the bin's initial thermal population $\bar{n}_{\rm th}$ is
\emph{resonantly} excited---it is not maintained by the drive and
therefore not replenished after a loss event.  Two processes contribute
at the same rate $\bar{n}_{\rm th}\,\kappa_1$: (i)~$T_1$ decay of an
existing thermal photon in the bin, and (ii)~absorption of a new
thermal photon from the bath during the gate.  Both transiently change
the bin photon number and thus the cross-Kerr phase imprinted on the
control, causing a bit-flip error.

At dilution-fridge base temperature $T_{\rm fridge} \approx 20$~mK
and bin frequency $\omega_b/2\pi \approx 5$~GHz, the Boltzmann
occupation is
\begin{equation}\label{eq:nth}
\bar{n}_{\rm th} = \frac{1}{e^{\hbar\omega_b/k_B T_{\rm fridge}} - 1}
\approx 6.2\times 10^{-6}.
\end{equation}
Conversely, the value $\bar{n}_{\rm th}=10^{-4}$ used in our
simulations corresponds to an effective mode temperature
\begin{equation}\label{eq:nth_teff}
T_{\rm eff} =
\frac{\hbar\omega_b}{k_B\ln(1+1/\bar{n}_{\rm th})}
\approx 26.1~{\rm mK}
\quad (\omega_b/2\pi = 5~{\rm GHz}).
\end{equation}
For comparison, $\bar{n}_{\rm th}=10^{-3}$ and $5\times 10^{-3}$
correspond to $T_{\rm eff}\approx 34.7$~mK and $45.2$~mK, respectively,
for the same mode frequency.
We therefore use $\bar{n}_{\rm th}=10^{-4}$ as a thermalization target
for a well-shielded high-$Q$ bin cavity, rather than as a value inferred
from the refrigerator temperature alone.  Direct residual-photon
measurements in superconducting 3D-cavity devices have bounded the
ambient cavity population at $2\times 10^{-4}$ with cavity attenuators
\cite{Wang2019CavityAttenuators} and with recent small-photon-number
metrology~\cite{Atalaya2024SmallPhoton}.  The assumed value
$10^{-4}$ is thus a factor-of-two improvement over the best direct
bounds we are aware of, not a claim of base-temperature Boltzmann
equilibration; residual excitations in superconducting circuits are
commonly limited by non-equilibrium mechanisms rather than the simple
Boltzmann floor~\cite{Serniak2018}.  This sets the thermal noise model value
$\bar{n}_{\rm th} = 10^{-4}$ used in the QEC analysis of
Fig.~\ref{main-fig:performance}; the leading control phase-flip contribution
scales linearly, $p_Z^{(c,{\rm th})}\approx\bar{n}_{\rm th}$, so results
at $\bar{n}_{\rm th}=10^{-3}$ or $5\times10^{-3}$ can be estimated by
raising this thermal floor by factors of 10 or 50, respectively.

The dashed lines in Fig.~\ref{fig:control_bin_T1} show the control
OFF bit-flip when the bin starts in a thermal state with
$\bar{n}_{\rm th} = 10^{-4}$ and experiences ongoing thermal heating.
At large $\bar{n}$, these curves saturate at a floor set by the
thermal contribution (dotted horizontal lines), rather than continuing
to decrease exponentially.  This thermal-limited floor is the
ultimate bit-flip limit for the control OFF channel.

\begin{figure}[t]
\includegraphics[width=\columnwidth]{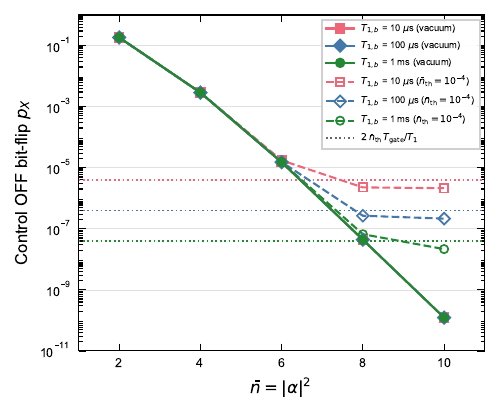}
\caption{\label{fig:control_bin_T1}
Control OFF bit-flip error versus $\bar{n} = |\alpha|^2$ for the
reduced control--bin system.
Solid lines: bin starts in vacuum with $T_1$ loss only---the
exponential suppression is maintained at all $T_1$ values,
confirming resilience to off-resonant bin photon loss.
Dashed lines: bin starts in thermal state with
$\bar{n}_{\rm th} = 10^{-4}$ and ongoing thermal heating from the
bath.
Dotted horizontal lines: thermal-limited bit-flip floor.
The thermal curves saturate at this floor at large~$\bar{n}$,
identifying bin thermal excitation as the ultimate limit on the
control OFF bit-flip.
Parameters: $\chi/2\pi = 4$~MHz, $\sigma = 0.12$,
$T_{\rm SWAP} = 100$~ns per elementary SWAP, $\chi$-echo.
}
\end{figure}

\section{$\chi$-echo: more details}
\label{sec:sm-chiecho}
In this section we detail the $\chi$-echo protocol that cancels the leading-order cross-Kerr phase accumulated during the gate.

\subsection{Two-pulse echo protocol}

\begin{enumerate}
\item \textbf{First half} ($0 \to T_{\rm SWAP}$): Evolve under $\hat{H}_1 = +\chi\,\hat{n}_c\hat{n}_b + g(t)(\hat{b}^\dagger\hat{a}_t + \hat{b}\,\hat{a}_t^\dagger)$ with beam-splitter pulse area $\pi/2$ (one elementary SWAP; Sec.~\ref{sec:swap-glossary}).

\item \textbf{Sign flip}: Reverse the cross-Kerr coupling, $\chi \to -\chi$ (potential hardware realizations are discussed in Sec.~\ref{sec:sm-hardware-realizations}).

\item \textbf{Second half} ($T_{\rm SWAP} \to 2\,T_{\rm SWAP}$): Evolve under $\hat{H}_2 = -\chi\,\hat{n}_c\hat{n}_b + g(t)(\hat{b}^\dagger\hat{a}_t + \hat{b}\,\hat{a}_t^\dagger)$ with pulse area $\pi/2$ (a second elementary SWAP, so that the two-pulse sequence realises SWAP$^2$).
\end{enumerate}

\noindent The total beam-splitter angle is $\theta = \pi/2 + \pi/2 = \pi$ (total pulse area $\pi$), realising the SWAP$^2$ that a hypothetical single continuous $\pi$-area pulse without an intervening $\chi$ sign flip would also realise.  The $\chi$-echo differs in that the cross-Kerr--induced phase now cancels to first order:
\begin{equation}
\varphi_{\rm echo} = \int_0^{T_{\rm SWAP}} \chi\,\bar{n}_b(t)\,dt
+ \int_{T_{\rm SWAP}}^{2\,T_{\rm SWAP}} (-\chi)\,\bar{n}_b(t)\,dt = 0
\end{equation}
when $\bar{n}_b(t)$ has the same time profile in both halves.  The two halves use identical pulse envelopes and the mean bin population is dominated by the far-detuned steady response, so the profiles nearly match; small differences from the initial-state mismatch at the midpoint contribute only at higher order in $\chi T_{\rm SWAP}$.  Note that this cancels only the deterministic phase from a bin occupancy that persists throughout the gate.  Assuming the bin equilibrates with the bath to $\bar n_{\rm th}$, both $T_1$ loss of an existing thermal photon and absorption of a new one from the bath occur at rate $\bar n_{\rm th}\kappa_1$; these \emph{stochastic} events insert or remove a bin photon at an arbitrary time within the two-pulse sequence, so the added photon contributes to only one half's integral and the cancellation fails on that fraction of the population (Sec.~\ref{sec:thermal_sensitivity}).

\subsection{Caveats and robustness}

The echo protocol assumes that the two pulse halves produce identical
beam-splitter dynamics modulo the sign of $\chi$.  Two effects can
modify this assumption:

\emph{Target ON bit-flip at $g < \chi$.}---For the target cat in the
ON branch, the echo can actually \emph{worsen} the bit-flip error
when $g_{\max} < \chi$.  In the SU(2) picture
(Sec.~\ref{sec:sm-su2}), the cross-Kerr detuning from residual
control excitations ($n_c = 1, 2, \ldots$) shifts the effective
beam-splitter rotation axis.  When the $\chi$-echo flips the sign of
$\chi$ at the midpoint, the two cSWAP halves rotate about
\emph{different} axes in the spin-$N/2$ Bloch sphere, so their
rotation errors compound rather than cancel.  For control states close
to but not exactly $|0\rangle$ (i.e., the $n_c \geq 1$ tail of the
displaced cat), this compounding produces a net target bit-flip that
is larger with echo than without.  When $g_{\max} \gg \chi$, the
cross-Kerr detuning becomes a small perturbation on the beam-splitter
rotation; the two axes are nearly identical and the echo actually
\emph{improves} the target bit-flip by refocusing the residual
rotation error.

\emph{$\chi$-echo magnitude mismatch.}---In practice, the sign flip
of $\chi$ may be imperfect: $\chi \to -(1-\delta)\chi$ with a
fractional mismatch $\delta$.  Simulations suggest that
$\delta \lesssim 5\%$ is sufficient for most of the $\chi$-echo's
phase-flip suppression benefits to remain.

\subsection{Alternative to $\chi$-echo: cycling condition}
\label{sec:sm-cycling-condition}

If implementing the $\chi$ sign flip is experimentally difficult, the
deterministic cross-Kerr phase from a constant bin thermal occupancy
can instead be nulled by choosing a total gate time that satisfies
the cycling condition $\chi\,T_{\rm gate} = 2\pi\,n$ (integer~$n$).
Under this condition, a bin thermal photon ($n_b = 1$ or any integer)
rotates the control in phase space by
$\chi\,n_b\,T_{\rm gate} = 2\pi\,n\,n_b$, which is a multiple of
$2\pi$ and therefore invisible.  This cancels the same
constant-$\bar n_{\rm th}$ contribution that the $\chi$-echo cancels,
so the cycling condition reaches the same
stochastic-bath-event-limited floor identified in
Sec.~\ref{sec:thermal_sensitivity} (neither protocol protects against
mid-gate photon loss or heating events).

Note that the cycling condition does \emph{not} null the phase from
the off-resonantly driven bin population, which is a coherent state
(not a Fock state) with continuously varying amplitude during the
pulse.  The trade-off of not using the $\chi$-echo is that this residual off-resonant bin population contributes to the control OFF bit-flip (Sec.~\ref{sec:sm-drag}), which must then be suppressed by pulse shaping alone.

When the cycling condition replaces the $\chi$-echo, the residual
ac~Stark phase on the target in the OFF branch
(Sec.~\ref{sec:sm-beamsplitter}) is no longer canceled by a sign flip
of~$\Delta$.  Instead, this phase must be corrected by the alternative
protocol described in Sec.~\ref{sec:sm-beamsplitter}: applying a small
detuning to the second cSWAP half-pulse in the ON branch so that both
branches acquire the same phase-space rotation, which is then absorbed
into the cat-qubit logical axis.

\section{Control phase-flip: local sources and displacement echo}
\label{sec:sm-dispecho}
This section addresses \emph{local} sources of control phase-flip---mechanisms acting on the control cat itself in the displaced basis $\{|0\rangle, |2\alpha_c\rangle\}$, where the two logical branches carry asymmetric mean photon numbers ($0$ and $4|\alpha_c|^2$).  In the undisplaced cat basis $|\pm\alpha_c\rangle$, any photon-number-conserving channel commutes with the photon-parity operator and therefore cannot connect the two parity eigenstates $|\mathcal C_\pm\rangle$; in the displaced frame this protection is lost, and the branch-asymmetric action of otherwise number-conserving processes maps directly onto a $Z$ error.  Two channels enter here: self-Kerr nonlinearity, which acts asymmetrically on the two branches and induces phase flips~\cite{Putterman2024cat}, and cavity pure dephasing, whose branch-dependent generator likewise contributes.  Both are suppressed by the displacement echo~\cite{Hann2024}: We derive how the echo converts the self-Kerr phase-flip scaling from $(K_a T)^2$ to $(K_a T)^4$, and quantify the residual $1/f$-dephasing contribution that survives the echo.
Non-local sources (bin thermal population and Kerr mismatch) are treated in the next section (Sec.~\ref{sec:sm-ctrl-pf}).

\subsection{Kerr without echo}
\label{sec:kerr_noecho}

The self-Kerr Hamiltonian
$\hat H_K = \tfrac{K_a}{2}\,\hat a^{\dagger 2}\hat a^2$
imprints a Fock-state--dependent phase $\phi_n = K_a n(n-1)t/2$, shearing
a coherent state into a crescent in phase space.  Over the gate time $T$
this induces a branch-differential rotation of the displaced cat.  The
scaling is set by the parity-odd combination $\hat H_+ - \hat H_-$ with
$\hat H_\pm \equiv \hat D^\dagger(\pm\alpha_c)\,\hat H_K\,\hat D(\pm\alpha_c)$
(defined below); a first-order expansion in $T$ of the parity expectation
value gives
\begin{equation}\label{eq:pZ_noecho}
\begin{aligned}
p_Z^{(c,\,\text{no echo})}
&\approx \tfrac{1}{2}(K_aT)^2\,
\langle\mathcal C_+|(\hat H_+-\hat H_-)^2|\mathcal C_+\rangle\\
&= 32(K_aT)^2\,|\alpha_c|^8 + O(|\alpha_c|^6),
\end{aligned}
\end{equation}
i.e.\ quadratic in $K_aT$ and \emph{octic} in $\alpha_c$.

\subsection{Displacement echo protocol}
\label{sec:dispecho_protocol}

The displacement echo~\cite{Hann2024} cancels the leading branch-differential
Kerr phase by exchanging which logical branch carries the large displacement
during the second half of the protocol.  In the displaced basis
$\{|0\rangle,|2\alpha_c\rangle\}$ this is accomplished by the sequence
$\hat D(\alpha_c) \to e^{-i\hat H_K\tau} \to \hat D(-2\alpha_c) \to e^{-i\hat H_K\tau} \to \hat D(\alpha_c)$
(Fig.~\ref{main-fig:chiecho}(e)), so the branch that begins at $|2\alpha_c\rangle$
is mapped to $|0\rangle$ at the midpoint (and vice versa).  Each branch spends
one segment near vacuum and one near large displacement, and the coherent
Kerr phase they accumulate is balanced.

\subsection{$(K_aT)^2 \to (K_aT)^4$ scaling derivation}
\label{sec:KT_scaling}

To compute the residual after cancellation, define the displaced Kerr
Hamiltonians
$\hat H_\pm \equiv \hat D^\dagger(\pm\alpha_c)\,\hat H_K\,\hat D(\pm\alpha_c)
= \tfrac{K_a}{2}\,A_\pm^{\dagger 2}A_\pm^2$
with $A_\pm \equiv \hat a\pm\alpha_c$, so that the echoed evolution
over $\tau = T/2$ per half is
$U_{\rm echo} = e^{-i\hat H_-\tau}\,e^{-i\hat H_+\tau}$.  The
photon-parity operator $\hat\Pi = e^{i\pi\hat a^\dagger\hat a}$ takes
$\hat a\pm\alpha_c \to -(\hat a\mp\alpha_c)$ and hence
$\hat H_+ \leftrightarrow \hat H_-$, so the parity-odd nested commutators
$\ell_{2m}$ ($m\geq 1$) in the Baker--Campbell--Hausdorff expansion of
$U_{\rm echo}$ vanish on $|\mathcal C_+\rangle$; the leading nonzero
contribution to $\langle\hat\Pi\rangle$ comes from squaring the
$\ell_2 = -\tau^2 C$ term with $C\equiv[\hat H_+,\hat H_-]$, since $C^2$
is parity-even.  Evaluating $C$ on $|\pm\alpha_c\rangle$ using
$A_\pm|\mp\alpha_c\rangle = 0$ gives
$C\,|\alpha_c\rangle = -2K_a^2\alpha_c^2\,A_-^{\dagger 2}\,|\alpha_c\rangle$
(and its parity partner).  Assembling
$\langle\mathcal C_+|C^2|\mathcal C_+\rangle$ with $\tau = T/2$ yields
\begin{equation}\label{eq:pZ_echo}
\boxed{
\begin{aligned}
p_Z^{(c,\,\text{echo})} \;\approx\;\;&
\tfrac{1}{8}(K_a T)^4\,|\alpha_c|^4\,\biggl[\tanh(|\alpha_c|^2)\\
&{}-\frac{8|\alpha_c|^2\bigl(|\alpha_c|^2-1\bigr)\,e^{-2|\alpha_c|^2}}{1+e^{-2|\alpha_c|^2}}\biggr].
\end{aligned}
}
\end{equation}
which for $|\alpha_c|^2\gg 1$ reduces to
$p_Z^{(c,\,\text{echo})} \approx \tfrac{1}{8}(K_aT)^4\,|\alpha_c|^4$.  Both this
expression and the no-echo formula~\eqref{eq:pZ_noecho} are leading-order
perturbative results, valid in the regime $K_aT\,|\alpha_c|^2 \ll 1$
where higher-order BCH terms are negligible; beyond this the exact
Bessel-series expansion of the Kerr channel must be used
[Sec.~\ref{sec:dispecho_numerics} validates the range of validity
numerically].  Compared to the unechoed $\sim 32(K_aT)^2\,|\alpha_c|^8$
scaling, the echo suppresses the phase-flip by many orders of magnitude
at the paper's operating point ($\chi/K\gtrsim 10^3$, $|\alpha_c|^2\lesssim 10$);
the numerical validation in Sec.~\ref{sec:dispecho_numerics} shows this
directly.

\subsection{Numerical validation of displacement echo scaling}
\label{sec:dispecho_numerics}

\begin{figure*}[t]
\centering
\includegraphics[width=\textwidth]{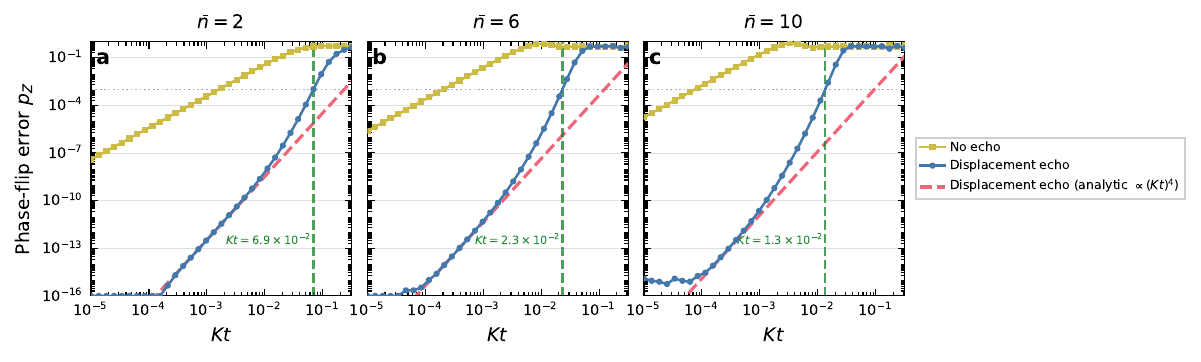}
\caption{%
Displacement echo phase-flip scaling.
Phase-flip error $p_Z$ vs.\ $KT$ for
$\bar{n} = 2,\,6,\,10$.
Blue: echoed; yellow: unechoed; red dashed: $(KT)^4$ analytic
[Eq.~\eqref{eq:pZ_echo}]; green dashed: $KT$ at which echoed
error crosses $10^{-3}$.
}
\label{fig:dispecho_scaling}
\end{figure*}

We validate the $(KT)^4$ scaling by numerically simulating the
single-mode displacement echo channel
\begin{equation}\label{eq:U_dispecho_channel}
U = \hat{D}(\alpha)\,
e^{-i\frac{K}{2}\hat{a}^{\dagger 2}\hat{a}^2\,\tau}\,
\hat{D}(-2\alpha)\,
e^{-i\frac{K}{2}\hat{a}^{\dagger 2}\hat{a}^2\,\tau}\,
\hat{D}(\alpha),
\end{equation}
applied to $|\mathcal{C}_+\rangle$, where $\tau = T/2$ is
half the total protocol time.  We sweep the dimensionless parameter
$KT$ (with $T$ fixed and $K$ varying) and extract the phase-flip
probability $p_Z = 1 - \langle\hat{J}_{++}\rangle$, where
$\hat{J}_{++}$ (defined in Sec.~\ref{sec:sm-error-defs}) is the even-parity cat projector.

Figure~\ref{fig:dispecho_scaling} shows the results for
$\bar{n} = |\alpha|^2 = 2,\,6,\,10$.  At small $KT$, the echoed
phase-flip (blue) follows the $(KT)^4$
approximation [Eq.~\eqref{eq:pZ_echo}] (red dashed) and is suppressed by
many orders of magnitude relative to the unechoed case (yellow).  At
larger $KT$, higher-order terms dominate and the perturbative formula
underestimates the true error, with the breakdown occurring at smaller
$KT$ for larger $\bar{n}$.

The green dashed lines mark the $KT$ value at which the echoed
phase-flip crosses $10^{-3}$.  For $\bar{n} = 10$, this threshold
corresponds to $K/2\pi \approx 4$~kHz (i.e., $\chi/K \approx 1000$
for $\chi/2\pi = 4$~MHz and a 400~ns total protocol).  This sets the
$\chi/K$ requirement for the displacement echo: Achieving
$p_Z^{(c)} \lesssim 10^{-3}$ at $\bar{n} = 10$ requires
$\chi/K \gtrsim 1000$.

\subsection{Pure dephasing on the displaced control}
\label{sec:sm-displaced-dephasing}

During the gate the control is displaced from $\{|\alpha\rangle,|{-}\alpha\rangle\}$ to $\{|2\alpha\rangle,|0\rangle\}$, so the two logical branches carry mean photon numbers $4|\alpha|^2$ and $0$.  Following the analysis of a related displaced-cat CR$_X$ gate in Ref.~\cite{Hann2024}, we model cavity pure dephasing as a stochastic frequency fluctuation $\hat H_{\rm deph}(t) = \delta(t)\,\hat a^\dagger\hat a$ with a $1/f$-dominated noise power spectrum $S_\delta(\omega) = A^2/|\omega|$ typical of flux-tunable superconducting circuits.  In the displaced frame the generator picks up a branch-dependent expectation value, and to leading order in Magnus expansion the parity-odd pieces average out under the mid-gate displacement flip $\hat{D}(-2\alpha)$; the surviving Z-error rate is set by the second cumulant of the displaced generator, giving
\begin{equation}\label{eq:pZ_dephasing_1f_echo}
p_{Z,\varphi}(\bar n) \approx 16\,\bar n^{2}\,(A\,T_{\mathrm{CX}})^{2}\log 2,
\end{equation}
i.e., quadratic in $\bar n$ and in $T_{\mathrm{CX}}$, with a $\log 2$ prefactor that is IR-cutoff independent because of the echo (cf.~Ref.~\cite{Hann2024}, Eqs.~(7) and~(20)).  We fix $A$ from the measured Ramsey coherence time $T_\varphi$ of a bare (undisplaced) cavity via the standard $1/f$ conversion $A = 1/(T_\varphi\sqrt{\log(1/(\omega_{\rm ir}T_\varphi))})$ with $\omega_{\rm ir}^{-1} \sim 1$~s taken as the characteristic experimental timescale for $1/f$ noise integration~\cite{Ithier2005}.  Throughout we assume $T_\varphi = 5\,T_1$, consistent with reported storage-cavity $T_1$ and Ramsey $T_2$ measurements~\cite{Putterman2024cqc, Putterman2024cat}, from which $T_\varphi/T_1 \approx 4$--$6$.  The resulting $p_{Z,\varphi}$ contribution enters the repetition-code noise model as a per-gate ancilla phase-flip probability (Sec.~\ref{sec:sm-noise-model}).

\section{Control phase-flip: non-local sources and Kerr-matching}
\label{sec:sm-ctrl-pf}
The previous section addressed the \emph{local} sources of control phase-flip---self-Kerr distortion of the control cat itself and cavity pure dephasing in the displaced frame---both suppressed by the displacement echo.
Here we focus on \emph{non-local} sources: mechanisms occurring outside the control cat that nonetheless cause a control phase-flip by modifying the SWAP$^2$ geometric phase or leaking which-path information.
Three mechanisms contribute: (i)~bin thermal population, (ii)~bin--target self-Kerr mismatch, and (iii)~cavity pure dephasing on the displaced control.

\subsection{Bin thermal population}
\label{sec:sm-ctrl-pf-thermal}

A thermal photon in the bin fundamentally alters the SWAP$^2$
geometric phase.  From the Fock-state result
(Sec.~\ref{sec:sm-su2}), SWAP$^2$ acting on a two-mode Fock state
gives
\begin{equation}
\text{SWAP}^2\,|m,n\rangle_{b,t} = (-1)^{m+n}\,|m,n\rangle_{b,t}.
\end{equation}
For the bin starting in vacuum ($m = 0$), each target Fock component
picks up $(-1)^n$, mapping $|\alpha_t\rangle \to |{-}\alpha_t\rangle$
as intended.  But if the bin contains a single thermal photon
($m = 1$), each component picks up $(-1)^{1+n} = -(-1)^n$---an
extra overall minus sign.

This extra sign has no effect on the target (it is a global phase on
the bin--target subsystem).  However, it is visible to the control.
In the ON branch, the control is displaced to $|0\rangle$ and the
CNOT unitary acts as
\begin{equation}
|0\rangle_c \otimes |m,\alpha\rangle_{b,t}
\;\xrightarrow{\;\text{SWAP}^2\;}
(-1)^m\,|0\rangle_c \otimes |m,{-}\alpha\rangle_{b,t}.
\end{equation}
With the control in a superposition (before displacement,
$|\alpha_c\rangle + |{-}\alpha_c\rangle$), the ON branch picks up
$(-1)^m$ while the OFF branch is unaffected (beam-splitter is
far off-resonant).  When $m = 1$, this produces a relative $-1$
between the two control branches---a control phase-flip.

Since $\bar{n}_{\rm th} \ll 1$, the bin thermal state is dominated
by the $m = 0$ and $m = 1$ components with probability
$p_1 \approx \bar{n}_{\rm th}$.  The $m = 1$ component flips the
control phase, giving a net control phase-flip probability
\begin{equation}\label{eq:ctrl_pZ_thermal}
p_Z^{(c,\,\text{th})} \approx \bar{n}_{\rm th}.
\end{equation}
For $\bar{n}_{\rm th} = 10^{-4}$, this gives
$p_Z^{(c,\,\text{th})} \approx 10^{-4}$, consistent with the
simulated thermal-limited floor in Fig.~\ref{main-fig:buildup}(d).

This error is fundamental: It arises from the parity structure of the
geometric phase and cannot be removed by the $\chi$-echo or
displacement echo (which address the OFF-branch cross-Kerr phase and
self-Kerr, respectively).  It can only be suppressed by reducing the
bin thermal occupation.

\subsection{Kerr-matching requirement ($K_b = K_t$)}
\label{sec:sm-kerr-mismatch}

Under Kerr-matching ($K_b = K_t$), the SWAP transfers the coherent
state between symmetric modes and both branches accumulate the same
nonlinear phase, so the self-Kerr contribution to control phase-flip
cancels to leading order.  If $K_b \neq K_t$
(Kerr-mismatch), the target cat
accumulates a different nonlinear phase in the bin than in its own
mode during the SWAP$^2$ cycle.  This asymmetry makes the ON and OFF
branches distinguishable (the target cat returns with a $K_b$-dependent
distortion in the ON branch but experiences only $K_t$ in the OFF
branch), leaking which-path information about the control state and
causing a control phase-flip error.

To quantify the tolerance, we perform a full three-mode simulation
(control + bin + target) with $\chi$-echo, DRAG, and displacement
echo, sweeping $K_b/K_t$ at fixed $K_t = \chi/1000$ and
$\bar{n}_t = 6$.

Figure~\ref{fig:kerr_mismatch} shows the control phase-flip
error~$p_Z^{(c)}$ versus the mismatch ratio $K_b/K_t$.  The curve
minimum sits at $p_Z^{(c)} = 1.1\times 10^{-4}$ near
$K_b/K_t \approx 1.25$---slightly offset from unity by higher-order
Kerr terms beyond the leading $K_b = K_t$ matching argument---while the
matched point $K_b/K_t = 1.0$ gives $p_Z^{(c)} = 1.6\times 10^{-4}$,
only marginally above the minimum.  Staying below the $p_Z < 10^{-3}$
threshold requires $K_b/K_t$ between approximately $0.2$ and $1.9$, a
comfortable tolerance given that bin and target modes can be designed
with similar geometry in superconducting circuits
(Sec.~\ref{sec:sm-hardware-realizations}).

\begin{figure}[t]
\centering
\includegraphics[width=\columnwidth]{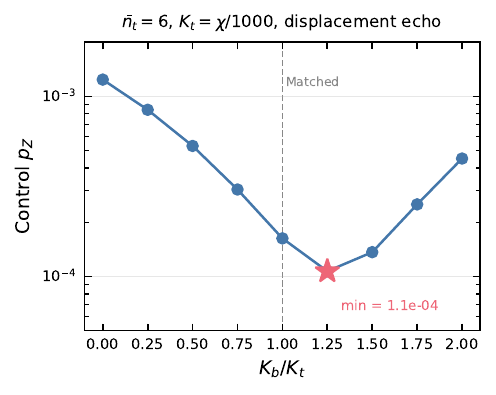}
\caption{\label{fig:kerr_mismatch}
Control phase-flip error versus bin-to-target self-Kerr ratio
$K_b/K_t$, with $K_t = \chi/1000$, $\bar{n}_t = 6$, and all
other parameters at the standard operating point ($T = 100$~ns,
$\sigma = 0.18$, $\chi/2\pi = 4$~MHz, displacement echo active).
Red star marks the minimum.  Dashed vertical: Kerr-matched condition
$K_b = K_t$.  The error remains below $10^{-3}$ for
$0.2 \lesssim K_b/K_t \lesssim 1.9$.
}
\end{figure}

\section{Idle-free control displacement echo}
\label{sec:sm-idle-free-echo}
The standard displacement echo (Sec.~\ref{sec:sm-dispecho}) requires
an idle period during the second half of the protocol, during which
the control sits at $|0\rangle$ or $|{-}2\alpha_c\rangle$ while no
beam-splitter drive is applied.  This idle segment doubles the total
gate time and increases photon-loss dephasing.  An alternative ``idle-free''
protocol eliminates this idle period by running a SWAP on the bin--target
pair in both halves of the echo, so that both control branches
accumulate equivalent nonlinear phases without dead time.

\subsection{Protocol}

Designing an idle-free variant is highly non-trivial: the bin must
never be significantly populated while the control simultaneously sits
at $|2\alpha_c\rangle$, or the cross-Kerr coupling $\chi\,\hat n_c\hat n_b$
generates control--bin entanglement that leaks which-path information
and creates a large control phase-flip.  The protocol below satisfies
this constraint by toggling $\chi$ on only during the two conditional
cSWAP$(\pi/2)$ steps (when the control is in $|0\rangle$ and bin
population is safe) and off during the two unconditional SWAP steps
(when both branches populate the bin symmetrically).

The idle-free control displacement echo proceeds as follows:
\begin{enumerate}
\item $D(\alpha_c)$: displace control from $\{|\alpha_c\rangle, |{-}\alpha_c\rangle\}$ to $\{|0\rangle, |2\alpha_c\rangle\}$.
\item Conditional SWAP$(\pi/2)$ with $\chi$ turned on: ON branch ($|0\rangle$) executes, OFF branch ($|2\alpha_c\rangle$) is detuned.
\item Unconditional SWAP$(\pi/2)$ with $\chi$ turned off: both branches execute.
\item $D(-2\alpha_c)$: swap roles; control goes from $\{|0\rangle, |2\alpha_c\rangle\}$ to $\{|{-}2\alpha_c\rangle, |0\rangle\}$.
\item Unconditional SWAP$(\pi/2)$ with $\chi$ turned off: both branches execute.
\item Conditional $-$SWAP$(\pi/2)$ with $\chi$ turned on: new ON branch ($|0\rangle$, formerly OFF) executes with opposite phase.
\item $D(\alpha_c)$: return control to $\{|\alpha_c\rangle, |{-}\alpha_c\rangle\}$.
\end{enumerate}

\noindent The $-$SWAP$(\pi/2)$ in step~6 denotes a beam-splitter pulse
with the coupling sign flipped ($g \to -g$), which physically
\emph{undoes} a standard cSWAP$(\pi/2)$.  The ON branch completes its
SWAP$^2$ in steps 2--3 (acquiring the $(-1)^n$ geometric phase), while
the OFF branch completes SWAP$(\pi/2)$ in step~3 followed by
$-$SWAP$(\pi/2)$ in step~6 (net identity).  Both branches have the bin
populated for exactly one SWAP duration.

If the SWAP time is limited by the achievable $\chi$, so that each
conditional cSWAP$(\pi/2)$ (steps~2 and~6) takes much longer than each
unconditional SWAP$(\pi/2)$ (steps~3 and~5), the total protocol time
is dominated by the two conditional steps and is approximately
$2\,T_{\rm cSWAP}$, versus $4\,T_{\rm cSWAP}$ for the standard
displacement echo with its idle segment.  In this regime the
idle-free protocol nearly halves the gate time, which
proportionally reduces photon-loss dephasing.

\subsection{Effect on control phase-flip channels}

\emph{Bin self-Kerr ($K_b$):} In the standard protocol, only the ON
branch populates the bin, so $K_b$ creates a branch-dependent phase
that survives even at the matched point $K_b = K_t$ (control
phase-flip channel of Sec.~\ref{sec:sm-kerr-mismatch}).  In the
idle-free protocol, both branches have identical bin population
duration $T_{\rm SWAP}$, eliminating this bin-population asymmetry
between branches and removing the corresponding contribution to
control phase-flip to leading order.

\emph{Bin thermal population:} The geometric phase argument of
Sec.~\ref{sec:sm-ctrl-pf-thermal} is unchanged: Odd bin thermal
photons flip the sign of the SWAP$^2$ phase in the ON branch while
the OFF branch (SWAP$\cdot(-$SWAP$)$ = identity) acquires no
geometric phase.  The relative $(-1)^m$ persists, giving
$p_Z^{(c)} \approx \bar{n}_{\rm th}$ as before.

\emph{Control self-Kerr ($K_a$):} In both protocols, the displacement
echo ensures each branch spends equal time at $|0\rangle$ and
$|2\alpha_c\rangle$, canceling the leading-order $K_a$ phase.  The
idle-free version achieves this without dead time.

\emph{Control dephasing:} Pure dephasing ($T_\varphi$) on the control
accumulates during the full gate time regardless of the protocol;
the idle-free echo reduces this by shortening the total gate time.

\subsection{Practical difficulties}

Despite these advantages, the idle-free protocol introduces two
practical complications that currently prevent its use:

\emph{(1) Control $|2\alpha_c\rangle$ fidelity after displacement back
to $|0\rangle$.}---During the gate, the control in $|2\alpha_c\rangle$
undergoes non-ideal evolution from entanglement with residual coherent
bin population (via cross-Kerr), self-Kerr distortion, and dephasing.
When this degraded state is displaced back to $|0\rangle$ by
$D(-2\alpha_c)$, it does not return to vacuum with high fidelity.
Crucially, most of these infidelity sources are \emph{not} suppressed
at large $|\alpha_c|$: The larger $|2\alpha_c|$ is, the smaller a
non-ideal rotation angle~$\theta$ can be tolerated, because even a
small $\theta$ maps to a displacement
$\sim 2\alpha_c\,\theta$ away from vacuum after $D(-2\alpha_c)$.

\emph{(2) Target bit-flip from imperfect control vacuum.}---If the
control state entering step~6 (after the second $D(-2\alpha_c)$
sequence and the intervening SWAPs) is not exactly $|0\rangle$ but
contains a small $|n_c\rangle$ component with $n_c \geq 1$, the
conditional $-$cSWAP$(\pi/2)$ acts on that component with the
cross-Kerr $\chi$ turned on: the $n_c$-quantum detunes the
beam-splitter by $\chi n_c$, producing an incomplete SWAP and a
target bit-flip error by the same physics as the thermal control
photon channel analyzed in Sec.~\ref{sec:sm-thermal-target-on}.  Note
that the unconditional SWAPs in steps~3 and~5 are performed with
$\chi$ turned off, so they are insensitive to $n_c$ at leading order;
this new vulnerability is confined to the second conditional step,
whereas the first conditional step (step~2) suffers only the usual
control-thermal-occupation channel of
Sec.~\ref{sec:sm-thermal-target-on} already present in the standard
protocol.

The above analysis assumes the cross-Kerr is sign-flipped by the
$\chi$-echo so that any bin phase acquired via $\chi\,n_c$ is
refocused.  If instead the cycling condition
$\chi\,t = 2\pi\,n$ (Sec.~\ref{sec:sm-cycling-condition}) is used to
null the constant-$\bar n_{\rm th}$ contribution, a related but
distinct timing constraint appears on the placement of the SWAP pulse
within the gate.  At the shortest
gate ($\chi\,t = 2\pi$), placing the SWAP pulse $g(t)$ in the middle
of the gate leaves bin photons to evolve for half the duration,
acquiring $\sim\!\pi$ of extra phase (for $n_c = 1$) and causing a
target bit-flip.  Two potential remedies both have significant costs:
\begin{itemize}
\item Use a very short $g(t)$ pulse at the start or end of the gate
  period, so that SWAP'd bin photons undergo a full $\chi\,t = 2\pi$
  rotation before returning to the target.  However, such a short
  pulse requires large peak amplitude $g_{\max}$, which increases the
  residual coherent bin population in the OFF branch and further
  degrades the control $|2\alpha_c\rangle$ fidelity.
\item Double the gate time to $\chi\,t = 4\pi$, so that a centrally
  placed SWAP produces bin photons that rotate by $2\pi$ (rather than
  $\pi$).  This eliminates the timing problem but doubles the gate
  duration---arguably no better than the standard protocol with its
  idle period.
\end{itemize}

\noindent For these reasons, we adopt the standard displacement echo
with idle period throughout this work.  The idle-free variant remains
an interesting direction for future optimization if the control
fidelity constraints can be relaxed.

\section{Longitudinal-coupling variant of the VCB mechanism}
\label{sec:sm-longitudinal}
The vacuum-conditional beam-splitter (VCB) mechanism developed in the main text is realised there through a cross-Kerr coupling $\chi\,\hat{a}_c^\dagger\hat{a}_c\,\hat{b}^\dagger\hat{b}$, which produces a branch-dependent bin frequency shift $\bar{\Delta} = \chi\langle\hat{n}_c\rangle$ that scales as $4\chi|\alpha_c|^2$ in the OFF branch and vanishes in the ON branch.  We note here that the same VCB gate can also be realised with a parametric longitudinal coupling
\begin{equation}\label{eq:H_longitudinal}
\hat{H}_{\rm long} = \lambda\,\bigl(\hat{a}_c + \hat{a}_c^\dagger\bigr)\,\hat{b}^\dagger\hat{b}
  + g(t)\bigl(\hat{b}^\dagger\hat{a}_t + \hat{b}\,\hat{a}_t^\dagger\bigr),
\end{equation}
in place of the cross-Kerr.  For a coherent state $|\alpha_c\rangle$, $\langle\hat{a}_c+\hat{a}_c^\dagger\rangle = 2\,\mathrm{Re}(\alpha_c)$, giving a bin detuning $\bar{\Delta} \approx 2\lambda\,\mathrm{Re}(\alpha_c)$: opposite in sign between the two logical branches $|\pm\alpha_c\rangle$ and continuously connected through vacuum.  In the displaced basis $\{|2\alpha_c\rangle, |0\rangle\}$ used by the gate, the OFF branch $|2\alpha_c\rangle$ acquires a detuning $\bar{\Delta}_{\rm OFF} = 4\lambda\,\mathrm{Re}(\alpha_c)$ while the ON branch $|0\rangle$ stays at resonance.

Compared with the cross-Kerr realisation, this variant has three qualitatively different features:
\begin{itemize}
\item \emph{Sign flip.} $\lambda$ is typically set by the amplitude and phase of a parametric pump; a $\pi$-shift of the pump phase inverts $\lambda \to -\lambda$ directly.  The $\chi$-echo protocol of Sec.~\ref{sec:sm-chiecho} then goes through unchanged, with pump-phase reversal in place of the flux-tuned cross-Kerr sign change discussed in Sec.~\ref{sec:sm-hardware-signflip}.
\item \emph{Branch-detuning scaling.} The bin detuning is now linear in $|\alpha_c|$ rather than quadratic, so the same off-resonance suppression of the OFF branch requires a proportionally larger $\lambda$ than $\chi$ at large cat size (roughly $\lambda \sim \chi\sqrt{\bar n}$ for equal detuning).
\item \emph{ON-branch approximation.} The control ON state (displaced vacuum $|0\rangle$) is not an exact eigenstate of $(\hat{a}_c + \hat{a}_c^\dagger)$; residual matrix elements of order $\sqrt{\langle\hat{n}_b(t)\rangle}$ couple the control to bin population during the gate, degrading the ON-branch bit-flip fidelity relative to the cross-Kerr case (where $|0\rangle$ is an exact eigenstate of $\hat{n}_c$).
\end{itemize}
The purpose of noting this variant is not to claim performance parity with the cross-Kerr scheme.  The main-text quantitative analysis is specific to the cross-Kerr Hamiltonian [Eq.~\eqref{main-eq:hamiltonian}], for which the ON branch is exact and the branch-detuning quadratic scaling is favorable.  Rather, the point is that the vacuum-conditional beam-splitter is the load-bearing physical idea, and its two ingredients---a branch-conditional bin detuning and a coherent beam-splitter swap---can be realised by more than one Hamiltonian.  A parametric longitudinal coupling is one such alternative that may prove useful in device families where either a large-$\chi$/low-$K$ cross-Kerr or a fast mid-gate $\chi$ sign change is harder to engineer than a pumped three-wave-mixing interaction of the form~\eqref{eq:H_longitudinal}~\cite{didierReadout, Potts2025}.  A full quantitative comparison, including the ON-branch fidelity trade-off and the drive-power budget for a specific coupler design, is a natural direction for follow-up studies.

\section{Potential hardware realizations}
\label{sec:sm-hardware-realizations}
The beam-splitter part of the proposal is comparatively mature:
parametric beam-splitter interactions between nearly linear bosonic
modes have reached MHz rates with high on/off ratios in bosonic-mode
platforms~\cite{Lu2023,Chapman2023}.  These experiments show that
fast, high-contrast exchange interactions between long-lived bosonic
modes are available.

The harder part is the cross-Kerr implementation.  In the operating
point of Fig.~\ref{main-fig:performance}, the gate uses
$\chi/2\pi \approx 4$~MHz while the parasitic self-Kerr of the cat and
bin modes must remain small.  From the displacement-echo scaling in
Sec.~\ref{sec:sm-dispecho}, keeping the echoed control self-Kerr
contribution below the $10^{-3}$ level at $\bar n \approx 10$ requires
$K/2\pi \lesssim 4$~kHz, or $\chi/K \sim 10^{3}$.  In addition, the
$\chi$-echo requires the sign of the control--bin cross-Kerr to be
reversible at the midpoint of the two-SWAP sequence.  These
requirements are demanding because cross-Kerr and self-Kerr terms
generally descend from the same nonlinear circuit element: a generic
shared Josephson nonlinearity produces both a mutual term
$\chi\,\hat n_c\hat n_b$ and local terms
$K_j\,\hat a_j^{\dagger 2}\hat a_j^2/2$.

\label{sec:sm-hardware-signflip}
One possible route is to start from principles from the quarton
coupler: use
opposite-sign Josephson nonlinearities so that self-Kerr terms cancel
while the cross-Kerr remains~\cite{Ye2021quarton, Arne}.  The recent quarton
coupler experiment demonstrates that this idea can produce very large
cross-Kerr rates, in the hundreds of MHz, together with strongly
suppressed Kerr in nearly linear modes~\cite{Ye2025quarton}.  Our gate
does not require such a large $\chi$; a few MHz is sufficient.  One
potential direction, which we outline only at the level of the leading
nonlinear terms, is to use a less aggressive flux-tunable galvanic
SQUID coupler~\cite{Kim2026squidCoupler,Kounalakis2018}, supplemented
by flux-tunable SQUIDs that cancel the self-Kerr induced on the two
modes being coupled.  The detailed operating point and mode
participation engineering would need to be worked out in future
device-specific studies.

For two modes with phase coordinates $\varphi_1$ and $\varphi_2$, the
relevant Josephson potential of such an idealized three-SQUID element
can be written as
\begin{equation}
\begin{aligned}
U(\varphi_1,\varphi_2)=&
-E_{J,c}(\Phi_c)\cos(\varphi_1-\varphi_2)\\
&-E_{J,1}(\Phi_1)\cos\varphi_1
-E_{J,2}(\Phi_2)\cos\varphi_2 ,
\end{aligned}
\end{equation}
where, for a symmetric SQUID,
$E_{J,\ell}(\Phi_\ell)=E_{J,\ell,\Sigma}\cos(\pi\Phi_\ell/\Phi_0)$
is flux tunable, including in sign.  Expanding to fourth order gives
\begin{align}
U^{(4)} = -\frac{1}{24}\big[
&(E_{J,1}+E_{J,c})\varphi_1^4
+(E_{J,2}+E_{J,c})\varphi_2^4 \nonumber\\
&-4E_{J,c}(\varphi_1^3\varphi_2+\varphi_1\varphi_2^3)
+6E_{J,c}\varphi_1^2\varphi_2^2
\big],
\end{align}
with all $E_{J,\ell}$ evaluated at their corresponding flux biases.  The
self-Kerr cancellation conditions are therefore
\begin{equation}
E_{J,1}(\Phi_1)+E_{J,c}(\Phi_c)\simeq0,\qquad
E_{J,2}(\Phi_2)+E_{J,c}(\Phi_c)\simeq0,
\end{equation}
up to the usual zero-point phase and participation-ratio factors of a
specific circuit mode.  Because all three SQUIDs are flux tunable,
these conditions can in principle be imposed at the chosen value of
$E_{J,c}(\Phi_c)$ and retuned when the sign of $E_{J,c}$ is reversed for the
$\chi$-echo.  After the self-Kerr terms are canceled, the only
diagonal resonant term left from the coupling potential is the
cross-Kerr generated by $\varphi_1^2\varphi_2^2$; the remaining
quartic monomials are non-number-conserving for nondegenerate modes
and can be minimized by choosing a sufficiently large mode detuning.

The purpose of this discussion is not to claim that the exact
operating point used in our simulations has already been demonstrated
in a single device.  Rather, it identifies the remaining hardware
requirement after the beam-splitter primitive is available and provides
a plausible starting path for its future demonstration: a flux-tunable
few-MHz cross-Kerr whose accompanying self-Kerr is canceled to the kHz
level.

\section{Pulse shaping}
\label{sec:sm-drag}
The OFF-branch control bit-flip error arises from transient population
of the bin mode by the off-resonant beam-splitter drive.  Here we
analyze how the pulse envelope $g(t)$ controls this leakage and
discuss the application of derivative removal by adiabatic gate (DRAG)
techniques~\cite{Motzoi2009} to the bin mode.

\subsection{Off-resonant bin excitation}

When the control is in the OFF state ($|2\alpha\rangle$), the
cross-Kerr interaction detunes the bin from the beam-splitter drive by
$\chi n_c$ per control Fock component, with an average detuning
$\bar{\Delta} = \chi\langle\hat{n}_c\rangle = 4\chi|\alpha_c|^2$.
Since the full dynamics is a superposition over control Fock states,
each component evolves independently at its own detuning
$\Delta = \chi n_c$; the analysis below applies to each such component.
For a time-dependent coupling $g(t)$,
the bin amplitude in the interaction picture obeys
(Sec.~\ref{sec:sm-beamsplitter})
\begin{equation}\label{eq:bin_amp}
i\dot{\tilde{b}} = g(t)\,e^{i\Delta t}\,\hat{a}_t,
\end{equation}
which for the bin initially in vacuum gives an instantaneous excitation
amplitude
\begin{equation}\label{eq:beta_t}
\beta(t) = -i\int_0^t g(t')\,e^{i\Delta t'}\,dt'\;\langle\hat{a}_t\rangle.
\end{equation}
The bin population remaining after the pulse is
\begin{equation}\label{eq:nbar_residual}
\bar{n}_b^{\rm final} = |\beta(T)|^2
\approx \frac{|\tilde{g}(\Delta)|^2}{\Delta^2}\,|\alpha_t|^2,
\end{equation}
where $\tilde{g}(\Delta) = \int_0^T g(t)\,e^{i\Delta t}\,dt$ is the
Fourier transform of the pulse envelope evaluated at the detuning
frequency.

\subsection{Adiabatic suppression via pulse shaping}

For a rectangular pulse ($g(t) = g_{\max}$ for $0 < t < T$), the
Fourier transform is a sinc function and the post-pulse population
oscillates as $\bar{n}_b^{\rm final} \sim (g_{\max}/\Delta)^2$, which
is suppressed only algebraically with $\Delta$.

A truncated Gaussian pulse of the form
\begin{equation}\label{eq:trunc_gaussian}
g(t) = g_{\max}\,\frac{\tilde g(t) - \tilde g(0)}{1 - \tilde g(0)},\qquad
\tilde g(t) = \exp\!\Bigl[-\frac{(t-T/2)^2}{2(\sigma T)^2}\Bigr],
\end{equation}
for $0\le t\le T$ (renormalized so that $g(0) = g(T) = 0$ exactly at the pulse boundaries)
has a Fourier transform that
falls off as $\exp(-\Delta^2\sigma^2 T^2/2)$, providing
\emph{exponential} suppression of the off-resonant excitation with
the ramp width~$\sigma$.

The Gaussian ramp only exponentially suppresses the off-resonant
excitation; it does not eliminate it.  A sharper pulse (smaller
$\sigma$) is less adiabatic and leaves more residual bin population
during the OFF branch, which we will show below is entangled with the
control through the cross-Kerr interaction and manifests as a control
bit-flip after the bin is traced out.  The precise mechanism, and the
role of DRAG in further suppressing this residual excitation, are
developed in the subsections that follow.  The gate-level simulations
in Fig.~\ref{main-fig:buildup} use $\sigma = 0.18$ with DRAG
($\gamma = 1.0$); see Sec.~\ref{sec:sm-gate-time} for the pulse-length
dependence.

\subsection{DRAG correction for the bin mode}

The DRAG technique~\cite{Motzoi2009} was originally developed to
suppress leakage to non-computational states in transmon qubits by
adding a derivative correction to the drive pulse.  For a two-level
subspace with detuning $\Delta$ and drive $g(t)$, the DRAG correction
adds an out-of-phase quadrature
\begin{equation}\label{eq:drag_correction}
g_{\rm DRAG}(t) = g(t) - i\,\frac{\dot{g}(t)}{\Delta},
\end{equation}
which cancels the leading-order non-adiabatic transition amplitude.

For the bin mode (a linear resonator rather than a transmon), the
same formalism applies~\cite{Hann2024} with the average detuning $\bar{\Delta} = 4\chi|\alpha_c|^2$: The DRAG
correction adds a derivative component to the beam-splitter drive
that destructively interferes with the off-resonant bin excitation.
The key difference from the transmon case is that the bin has no
anharmonicity, so higher-order DRAG corrections (which account for
transmon nonlinearity) are unnecessary.

The DRAG detuning parameter $\gamma$ controls the effective
detuning used in the derivative correction:
$\Delta_{\rm DRAG} = \gamma \times 4\chi|\alpha_c|^2$.
We first study the DRAG correction in a simplified two-mode model
(control + bin only): Since the OFF-branch target amplitude is
approximately constant, the beam-splitter
$g(\hat{b}^\dagger\hat{a}_t + \hat{b}\,\hat{a}_t^\dagger)$ reduces to
a classical drive $g\alpha_t(\hat{b} + \hat{b}^\dagger)$ on the bin,
making the dynamics tractable within the control--bin Hilbert space.
We caution that the classical treatment of the target as a fixed
$\alpha_t$ is only qualitatively correct: In the full three-mode
simulations of the following subsections, the DRAG quadrature
drives the bin which in turn acts back on the quantum target, an
effect entirely absent here.  The optimal $\gamma$ values found below
therefore do not transfer quantitatively to the full gate.

Figure~\ref{fig:drag_gamma} shows the post-gate bin population and
control bit-flip error as a function of $\gamma$ for several cat
amplitudes, using this reduced model with $T_{\rm pulse} = 75$~ns
per half ($150$~ns total) and $\sigma = 0.1 T_{\rm pulse}$.  The data
show that the residual bin population and the
$|2\alpha_c\rangle$-branch control bit-flip track each other closely
across the swept parameter range.  This confirms that the residual
bin population is an excellent proxy for the control bit-flip: the
bin becomes entangled with the control's $|2\alpha_c\rangle$
component through the cross-Kerr, so any excitation left in the bin
at the end of the gate decoheres the control when the bin is traced
out.

\begin{figure}[t]
\includegraphics[width=\columnwidth]{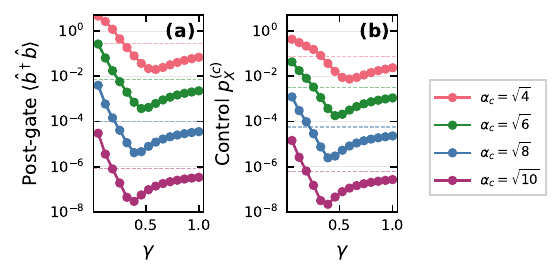}
\caption{\label{fig:drag_gamma}
DRAG detuning fraction $\gamma$ sweep.
(a)~Final bin population $\langle \hat{b}^\dagger \hat{b} \rangle$ and
(b)~control bit-flip error versus $\gamma$ for
$\alpha_c = \sqrt{4}$, $\sqrt{6}$, $\sqrt{8}$, $\sqrt{10}$.
The control bit-flip closely tracks the residual bin population,
confirming that leftover bin excitation is the source of the
$|2\alpha_c\rangle$-branch bit-flip through cross-Kerr entanglement.
Dashed horizontal lines indicate the no-DRAG baseline for each $\alpha_c$.
The optimal $\gamma$ decreases with increasing $\alpha_c$, converging
to $\gamma^* \approx 0.4$ for $|\alpha_c|^2 \geq 8$.
Parameters: $\chi/2\pi = 4$~MHz, $T_{\rm pulse} = 75$~ns,
$\sigma = 0.1\,T_{\rm pulse}$.
}
\end{figure}

At each $\alpha_c$, there is a clear optimal $\gamma^*$ that minimizes
the bit-flip error: Too small a $\gamma$ over-corrects (the DRAG
quadrature $\dot{g}/\Delta_{\rm DRAG}$ becomes large and drives the
bin harder than the original leakage), while too large a $\gamma$
under-corrects.
The optimal $\gamma^*$ decreases monotonically with $|\alpha_c|^2$,
from $\gamma^* \approx 0.6$ at $|\alpha_c|^2 = 4$ to
$\gamma^* \approx 0.4$ at $|\alpha_c|^2 \geq 8$,
reflecting the increasingly perturbative nature of the off-resonant
drive at large detuning.
At the optimum, DRAG improves the control bit-flip by $5$--$8\times$
across the range $|\alpha_c|^2 = 4$--$10$.

\subsection{Full DRAG with detuning correction}

Beyond the derivative quadrature correction of
Eq.~\eqref{eq:drag_correction}, the Motzoi framework~\cite{Motzoi2013}
provides two additional corrections that compensate the AC Stark shift
and rotation-angle distortion induced by the DRAG drive.  For a drive
with Rabi frequency $\Omega(t)$ targeting a transition at detuning
$\Delta$, the combined corrections are
\begin{align}
\mathrm{Im}\{\Omega\} &= -\frac{\dot{\Omega}_0}{\Delta},
  \label{eq:drag_q}\\
\delta(t) &= -2\,\Omega_0(t)\,\tan\!\Bigl(\frac{2\,\Omega_0(t)}{\Delta}\Bigr),
  \label{eq:drag_det}\\
\Omega_0 &\to r\,\Omega_0,\quad
\int_0^T r\,\Omega_0\,\sec\!\Bigl(\frac{2\,r\,\Omega_0}{\Delta}\Bigr)dt
  = \frac{\pi}{2},
  \label{eq:drag_amp}
\end{align}
where Eq.~\eqref{eq:drag_q} is the derivative quadrature correction,
Eq.~\eqref{eq:drag_det} is the detuning correction that compensates
the AC Stark shift on the resonant sector, and
Eq.~\eqref{eq:drag_amp} is the amplitude rescaling that preserves the
rotation angle.

The effectiveness of the detuning correction depends critically on the
\emph{speed ratio} $\varepsilon_{\rm peak}/|\Delta|$, where
$\varepsilon_{\rm peak}$ is the peak Rabi frequency driving the bin;
in our beam-splitter setting the effective bin drive from the
coherent target reduces to
$\varepsilon_{\rm peak}\sim g_{\max}|\alpha_t|$.
When the speed ratio exceeds ${\sim}0.2$, the $\tan$ function in
Eq.~\eqref{eq:drag_det} exits the perturbative regime and the
correction becomes comparable to or larger than the detuning itself,
degrading performance.

For our $\chi$-echo gate, the speed ratio is controlled by the
pulse width~$\sigma$: Narrower pulses concentrate more energy at the
peak, increasing $\varepsilon_{\rm peak}$ for a given pulse area.
Figure~\ref{fig:drag_sigma_gamma} shows the bit-flip improvement from
the full DRAG correction [Eqs.~\eqref{eq:drag_q}--\eqref{eq:drag_amp}]
as a function of both $\sigma/T$ and $\gamma$, obtained from the
full three-mode (control + bin + target) $\chi$-echo simulation at
$\bar{n} = 6$ and $T = 75$\,ns.
The full simulation captures an effect absent from the reduced model
of Fig.~\ref{fig:drag_gamma}: The DRAG quadrature drive on the bin
acts back on the target through the beam-splitter coupling, which the
reduced model (classical target) ignores.

\begin{figure}[t]
\includegraphics[width=\columnwidth]{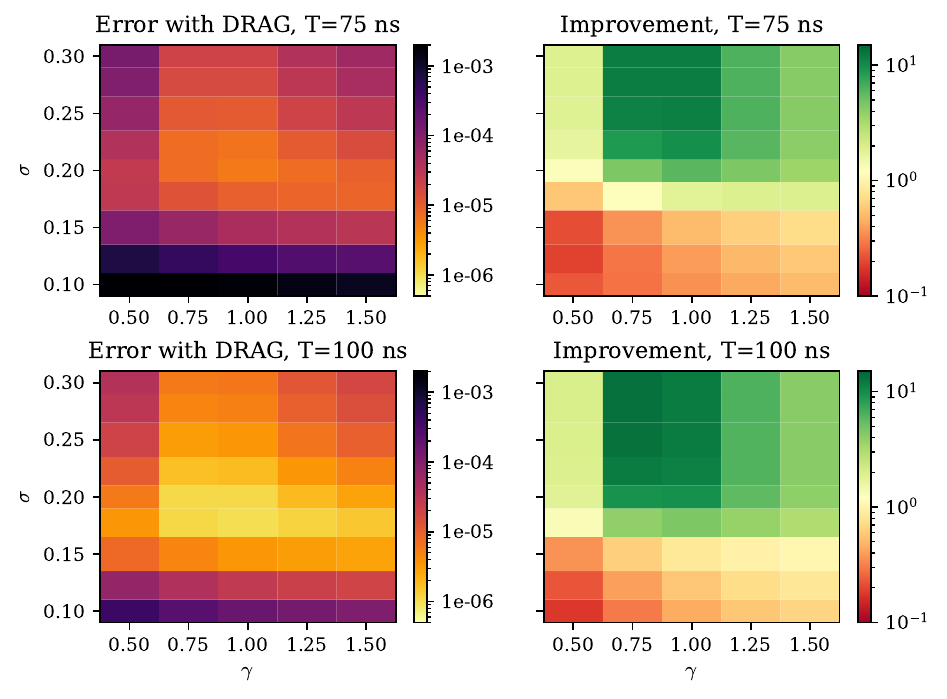}
\caption{\label{fig:drag_sigma_gamma}
DRAG pulse optimization in the $(\sigma, \gamma)$ parameter space.
Left column: absolute control bit-flip error (OFF branch) with full DRAG
[Eqs.~\eqref{eq:drag_q}--\eqref{eq:drag_amp}] at $T = 75$\,ns (top)
and $T = 100$\,ns (bottom).
Right column: improvement factor relative to the no-DRAG baseline at the
same~$\sigma$; values ${>}1$ (green) indicate improvement.
The optimal DRAG operating point is $\sigma \approx 0.18$--$0.20$,
$\gamma \approx 0.75$--$1.0$.  At $\sigma \lesssim 0.15$, DRAG
degrades performance (red region).
Parameters: $\chi/2\pi = 4$\,MHz, $\bar{n} = 6$.
}
\end{figure}

The results reveal a fundamental trade-off between the pulse width
and the effectiveness of the DRAG detuning correction:
\begin{enumerate}
\item \emph{Narrow pulses} ($\sigma \lesssim 0.15$): The speed
  ratio $\varepsilon_{\rm peak}/|\Delta|$ exceeds ${\sim}0.2$,
  causing the detuning correction to overcorrect.  DRAG becomes
  harmful for the control bit-flip (red region in
  Fig.~\ref{fig:drag_sigma_gamma}).
\item \emph{Broad pulses} ($\sigma \gtrsim 0.18$): The speed ratio
  is low enough for the detuning correction to remain perturbative.
  The full DRAG correction improves the control bit-flip relative
  to the no-DRAG baseline at the same~$\sigma$.
\end{enumerate}

However, the no-DRAG baseline itself depends on~$\sigma$: It reaches
a minimum at $\sigma \approx 0.15$--$0.18$ (where the Gaussian
roll-off optimally suppresses off-resonant leakage) and degrades
at larger~$\sigma$ due to increased post-gate bin population from the
broader pulse.  The \emph{absolute} improvement from DRAG, comparing
the best DRAG operating point ($\sigma \approx 0.20$, $\gamma = 1.0$)
against the best no-DRAG baseline ($\sigma \approx 0.15$), is
approximately $3\times$ at $\bar{n} = 6$, consistent across
both pulse lengths ($T = 75$ and $100$\,ns).

The detuning fraction $\gamma \approx 1$ is optimal at all tested cat
sizes ($\bar{n} = 4$--$10$), corresponding to a DRAG detuning
$\Delta_{\rm DRAG} = 4\chi|\alpha_c|^2$ that matches the average
cross-Kerr detuning $\chi\langle\hat{n}_c\rangle$.

\subsection{Gate time dependence of error channels}
\label{sec:sm-gate-time}

The preceding analysis focused on a single pulse length; here we
examine how the gate errors evolve as the elementary SWAP duration
$T_{\rm SWAP}$ is varied from 25 to 100~ns (50--200~ns total gate time).
For each~$T$, we independently optimize the DRAG parameters
$(\sigma, \gamma)$ via a two-dimensional sweep at $\bar{n} = 6$,
then evaluate all error channels across $\bar{n} = 2$--$10$.

Figure~\ref{fig:errors_vs_T} shows the resulting bit-flip and
phase-flip errors for $T = 25$, $50$, $75$, and $100$~ns.  Several
features are notable:
\begin{enumerate}
\item \emph{Control cat errors improve rapidly with $T$.}
  Since self-Kerr and $T_1$ are excluded here, the control OFF
  bit-flip and control phase-flip are both coherent errors from
  imperfect adiabaticity.  Both arise from the same mechanism: the
  transient off-resonant bin excitation becomes entangled with the
  control cat via the cross-Kerr interaction, and if these bin photons
  remain at the end of the gate, tracing out the bin decoheres the
  control's $|2\alpha_c\rangle$ branch relative to $|0\rangle$.  The
  transient bin population is set by the adiabaticity
  ratio $g_{\rm max}/\bar{\Delta}$: longer pulses require a smaller
  peak drive amplitude for the same pulse area, suppressing the
  entanglement.  At $\bar{n}=10$, the control phase-flip drops from
  $2\times 10^{-2}$ ($T=25$~ns) to $10^{-4}$ ($T=100$~ns); at
  $\bar{n}=6$, the control OFF bit-flip drops from $1.2\times 10^{-3}$
  to $10^{-6}$ over the same range.
\item \emph{Target bit-flips are well preserved across all $T$.}
  The target ON and OFF bit-flip errors are near or below the idling
  bound $\sqrt{\bar{n}}\,e^{-2\bar{n}}$ at all pulse lengths,
  confirming that the SWAP$^2$ mechanism faithfully preserves the
  target cat-qubit bit-flip suppression.
\item \emph{DRAG effectiveness degrades at short $T$.}  The optimal
  DRAG parameters shift with pulse length: $\gamma = 1.0$ is optimal
  for $T \geq 50$~ns, while at $T = 25$~ns the optimal $\gamma$
  drops to $0.5$ and the improvement over baseline reduces from
  ${\sim}3\times$ to ${\sim}2\times$.  This reflects the breakdown
  of the perturbative DRAG regime when the speed ratio exceeds
  ${\sim}0.3$.
\end{enumerate}

\begin{figure*}[t]
\includegraphics[width=\textwidth]{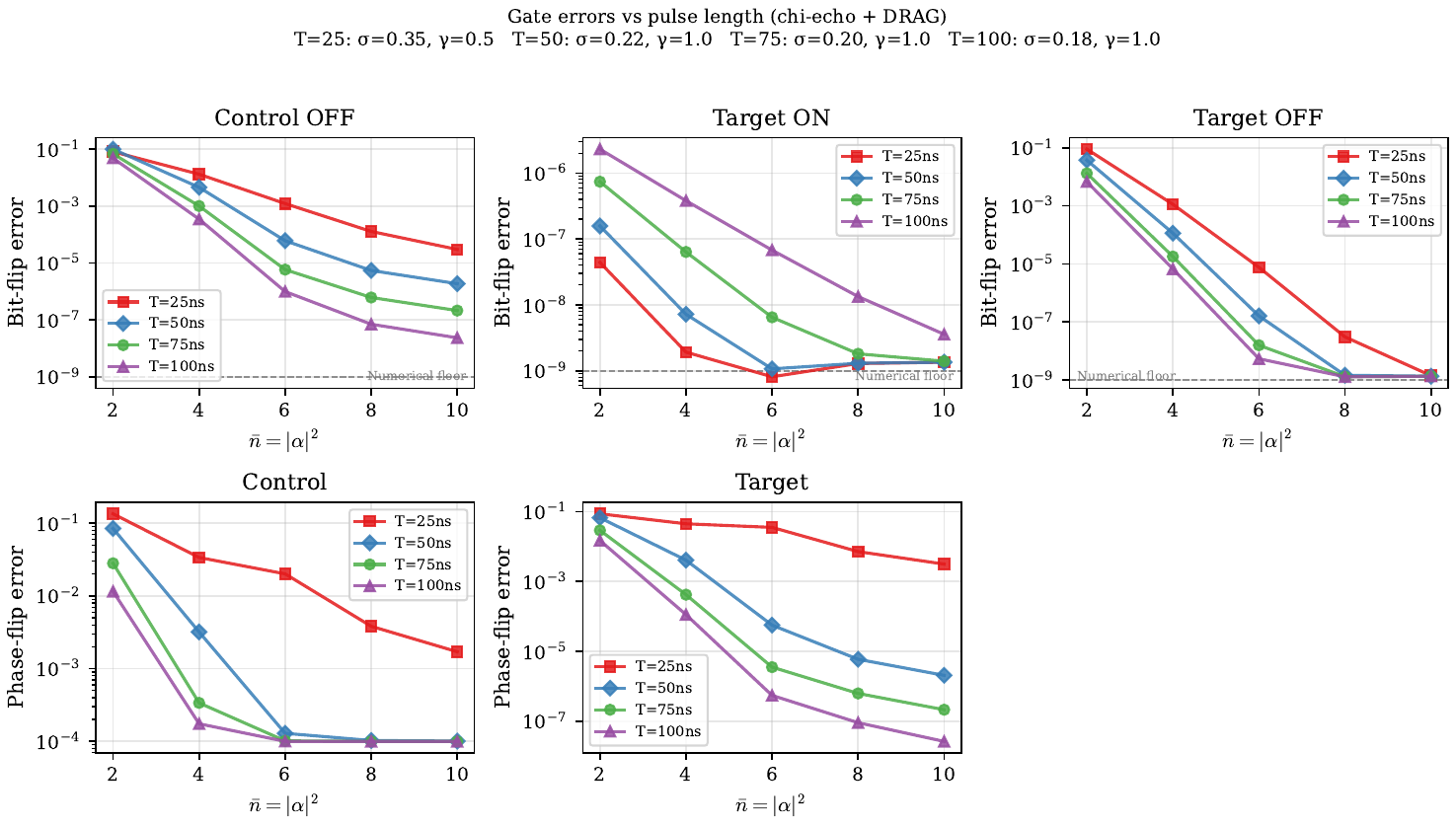}
\caption{\label{fig:errors_vs_T}
Gate errors versus $\bar{n} = |\alpha|^2$ at four pulse lengths,
each with independently optimized DRAG parameters.
Top row: bit-flip errors for the three nontrivial channels.
Bottom row: phase-flip errors for control and target.
Unlike the buildup panels of Fig.~\ref{main-fig:buildup}, only the
coherent gate dynamics are simulated here---self-Kerr ($K_a=K_b=0$),
thermal bin population, and single-photon loss ($T_1$) are excluded
so that only the pulse-length dependence of the coherent error
channels is isolated.  These excluded channels are the dominant
contributions to the target ON bit-flip, control phase-flip, and
target phase-flip in a realistic device, so the levels in the
corresponding panels are lower bounds set by pulse-shaping alone.
The control OFF bit-flip (top left) is the error that distinguishes
the two operating regimes: It is negligible at $T \geq 75$~ns but
becomes significant at shorter pulse lengths.
DRAG parameters (optimized at $\bar{n}=6$):
$T=25$~ns ($\sigma=0.35$, $\gamma=0.5$),
$T=50$~ns ($\sigma=0.22$, $\gamma=1.0$),
$T=75$~ns ($\sigma=0.20$, $\gamma=1.0$),
$T=100$~ns ($\sigma=0.18$, $\gamma=1.0$).
$\chi/2\pi = 4$~MHz throughout.
}
\end{figure*}

These results motivate two distinct operating points for different
code architectures:
\begin{itemize}
\item \emph{Syndrome extraction} ($T = 50$~ns):  In a repetition code,
  the control cat is the measurement ancilla, whose bit-flip errors do
  not propagate to the data qubits (Sec.~\ref{sec:sm-noise-model}).  At $T = 50$~ns and $\bar{n}=6$,
  both phase-flip errors are below $1.3 \times 10^{-4}$ and the target
  bit-flip is well preserved, while the control OFF bit-flip
  ($6\times 10^{-5}$) is irrelevant.  The shorter gate time also
  reduces $T_1$-induced idle dephasing on the data qubits during
  syndrome extraction, which is the dominant data-cat phase-flip
  channel identified in the buildup panels of
  Fig.~\ref{main-fig:buildup}.
\item \emph{Transversal logical CNOT} ($T = 100$~ns):  For a
  transversal CNOT between two repetition-code blocks, bit-flip errors
  on both control and target propagate logically.  The longer pulse
  suppresses the control OFF bit-flip to ${\sim}10^{-7}$ at
  $\bar{n} = 8$ and further to a few $10^{-8}$ at $\bar{n} = 10$,
  ensuring that all error channels remain well below the code
  threshold.
\end{itemize}

For the QEC simulations of Fig.~\ref{main-fig:performance}, the gate errors
are evaluated at the syndrome-extraction operating point ($T = 50$~ns)
with the ancilla cat fixed at $\bar{n}_a = 8$ and the data-cat
$\bar{n}$ swept from 2 to~10.
The choice $\bar{n}_a = 8$ (rather than $\bar{n}_a = 6$) is motivated by the
target OFF-branch bit-flip: In the OFF branch, the ancilla control state
$|2\alpha_a\rangle$ has a Fock-space tail at low photon numbers
($P(n\leq 3) \sim e^{-4\bar{n}_a}(4\bar{n}_a)^3/6$) that makes the
beam-splitter weakly resonant, driving a small unwanted SWAP of the target.
At $\bar{n}_a = 6$ ($|2\alpha_a|^2 = 24$), this tail gives
$P(n=3) \approx 10^{-7}$, limiting the target OFF bit-flip to
${\sim}1.4\times 10^{-7}$---comparable to the photon-loss bit-flip floor.
Increasing to $\bar{n}_a = 8$ ($|2\alpha_a|^2 = 32$) suppresses
$P(n=3)$ by ${\sim}10^3$, pushing the target OFF bit-flip well below
the numerical simulation floor.
Figure~\ref{fig:qec_gate_input} shows the resulting per-gate bit-flip
and phase-flip probabilities used as inputs to the circuit-level noise model.

\begin{figure}[!htbp]
\includegraphics[width=\columnwidth]{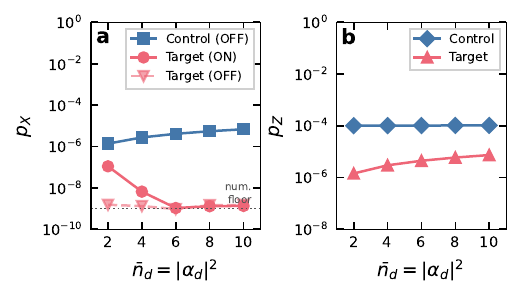}
\caption{\label{fig:qec_gate_input}
Per-gate error probabilities at the syndrome-extraction operating point
($T = 50$~ns, $\sigma = 0.22$, DRAG $\gamma = 1.0$) with ancilla fixed at
$\bar{n}_a = 8$.
These are coherent (Schr\"{o}dinger) simulations with no dissipation;
photon-loss dephasing ($\bar{n}\,T_{\rm CX}/T_1$) and photon-loss bit-flip
($\bar{n}\,e^{-2\bar{n}}\,T_{\rm CX}/T_1$) are added analytically in the
QEC noise model of Fig.~\ref{main-fig:performance}.
(a)~Bit-flip errors (OFF branch): control (blue squares) and target
(red circles: ON; red triangles: OFF).
(b)~Phase-flip errors: control (blue diamonds) and target (red triangles).
}
\end{figure}

\FloatBarrier

\section{Repetition code QEC: more details}
\label{sec:sm-simcnot}
In the repetition cat code~\cite{Guillaud2019, Chamberland2022}, each syndrome extraction round applies $2(d{-}1)$ CNOT gates between the $d$ data cats and $d{-}1$ ancilla cats: Each ancilla couples to its two neighboring data cats via one CNOT each, arranged in the standard two parallel steps.  This yields the round time $2\,T_{\rm CX} + T_{\rm meas}$, independent of code distance, with the ancilla remaining in the displaced computational basis $\{|0\rangle,|2\alpha_c\rangle\}$ throughout both steps, and with the displacement echo and two-photon stabilization on the ancilla applied only after both steps complete.  Deferring the stabilization until both CNOTs are done is crucial: If stabilization were re-engaged between the two CNOTs it would collapse the ancilla in the $\{|\pm\alpha_c\rangle\}$ basis, so any ancilla $|2\alpha_c\rangle$-branch rotation acquired during the first CNOT would be projected into an ancilla bit-flip that then propagates into the second CNOT and onto its data cat.

\subsection{Parallel beam-splitter drives}

Consider a distance-$d$ repetition code with data cats
$\hat{a}_{d,1},\ldots,\hat{a}_{d,d}$ and ancilla cats
$\hat{a}_{a,1},\ldots,\hat{a}_{a,d-1}$.  In the standard layout,
ancilla $k$ couples to data cats $k$ and $k{+}1$ via bin modes
$\hat{b}_{k,L}$ and $\hat{b}_{k,R}$.  During each CNOT step the
Hamiltonian is
\begin{equation}\label{eq:Hsim}
\hat{H}_{\rm sim} = \sum_{k=1}^{d-1}\Bigl[
  \chi_k\,\hat{n}_{a,k}\,\hat{n}_{b,k}
  + g_k(t)\bigl(\hat{b}_k^\dagger\hat{a}_{d,j(k)}
  + \hat{b}_k\,\hat{a}_{d,j(k)}^\dagger\bigr)
\Bigr],
\end{equation}
where the index $j(k)$ selects the data cat being probed in the
current step (either $k$ or $k{+}1$).  Since each bin mode
$\hat{b}_k$ couples to only \emph{one} data cat at a time and the
cross-Kerr terms involve only the local ancilla--bin pair, the
individual CNOT Hamiltonians commute:
$[\hat{H}_k, \hat{H}_{k'}] = 0$ for $k \neq k'$.  All $d{-}1$ CNOTs
therefore execute independently in parallel.

\subsection{Two-step round}

A complete syndrome extraction round proceeds in two steps
(Fig.~\ref{main-fig:performance}(a)):

\begin{enumerate}
\item \textbf{Step~1:} Each ancilla $k$ performs a CNOT with data cat
  $k$ (the ``left'' neighbor).  All $d{-}1$ CNOTs run simultaneously.
\item \textbf{Step~2:} Each ancilla $k$ performs a CNOT with data cat
  $k{+}1$ (the ``right'' neighbor).  Again all $d{-}1$ CNOTs in
  parallel.
\end{enumerate}

\noindent Each CNOT consists of a $\chi$-echo cSWAP$^2$ (two elementary SWAPs, duration $2\,T_{\rm SWAP}$; Sec.~\ref{sec:swap-glossary}).  Since the ancilla remains in its computational basis $\{|0\rangle, |2\alpha\rangle\}$ after each cSWAP$^2$, the two steps can execute back-to-back without an intermediate displacement echo.  A single displacement echo (duration $4\,T_{\rm SWAP}$) follows both steps to refocus the ancilla self-Kerr accumulated over the full $4\,T_{\rm SWAP}$ gate period.  The time structure is therefore
\begin{equation}
\underbrace{[0,\,2\,T_{\rm SWAP}]}_{\text{Step 1 cSWAP}^2}
\;\underbrace{[2\,T_{\rm SWAP},\,4\,T_{\rm SWAP}]}_{\text{Step 2 cSWAP}^2}
\;\underbrace{[4\,T_{\rm SWAP},\,8\,T_{\rm SWAP}]}_{\text{displacement echo}}\,.
\end{equation}
The total round time is $8\,T_{\rm SWAP} + T_{\rm meas}$ and the effective per-gate time is $T_{\rm CX} = 4\,T_{\rm SWAP}$.
For $T_{\rm SWAP} = 50$~ns and $T_{\rm meas} = 200$~ns subsuming ancilla readout and reset, this gives $T_{\rm CX} = 200$~ns and a round time of~$600$~ns.

After both steps, the ancilla state encodes the joint
parity $\hat{Z}_{d,k}\hat{Z}_{d,k+1}$ of the two neighboring data
cats, which is read out by measuring the ancilla in the $Z$ basis.

\subsection{Effect on logical error rate}

The dominant per-round physical error is data-cat phase-flip from photon-loss dephasing, which scales as $p_Z \propto |\alpha|^2\,T_{\rm round}/T_1$ per data qubit per round.  With the two-layer schedule above, $T_{\rm round} = 2\,T_{\rm CX} + T_{\rm meas}$ is independent of code distance $d$.  Combined with the gate-intrinsic phase-flip $p_Z^{\rm gate}$, the per-round physical phase-flip probability is
\begin{equation}\label{eq:pZ_round}
p_Z^{\rm round} \approx \frac{|\alpha|^2 (2\,T_{\rm CX} + T_{\rm meas})}{T_1}
+ p_Z^{\rm gate},
\end{equation}
which is decoded by MWPM to give a logical $Z$ rate suppressed as $\sim (p_Z^{\rm round})^{d/2}$ in the sub-threshold regime (see Fig.~\ref{main-fig:performance}).

\subsection{Circuit-level noise model}
\label{sec:sm-noise-model}

\emph{Setup and parameters.}
The logical error rate results in
Fig.~\ref{main-fig:performance}(b,c) come from injecting a
circuit-level Pauli noise model into a distance-$d$ repetition code
and decoding with pymatching.  Throughout, we take
$T_1 = 1$~ms, $T_{\mathrm{CX}} = 200$~ns, $T_{\mathrm{meas}} = 200$~ns,
$K_a/2\pi = 10$~kHz, and measurement error $p_{\rm meas} = 10^{-3}$;
ancillas run at fixed $\bar{n}_a = 8$ and are measured/reset after each
syndrome round.  The CNOT noise channel is a \emph{product} of
independent single-qubit Pauli channels on the data and ancilla
sides---no correlated two-qubit errors are injected---with rates
taken from the gate-level simulations of
Fig.~\ref{fig:qec_gate_input}.

\emph{Per-CNOT rates.}
The data-cat bit-flip rate $p_X^{(t)}(\bar{n})$ is obtained by
log-space interpolation of the coherent-gate simulation.  The
coherent gate-induced target bit-flips (both ON and OFF branches)
are at the numerical simulation floor (${\sim}10^{-9}$) for
$\bar{n}\geq 4$ at $\bar{n}_a = 8$; the logical $X$ error is
therefore set by the photon-loss formula
$d\,\kappa\,\bar{n}\,(2T_{\rm CX} + T_{\rm meas})\,e^{-2\bar{n}}$
rather than the interpolated $p_X^{(t)}$.
The ancilla bit-flip rate is set analytically from the photon-loss
channel,
\begin{equation*}
p_X^{(c)} = (\bar{n}_a\,T_{\rm CX}/T_1)\,e^{-2\bar{n}_a};
\end{equation*}
the coherent-gate ancilla bit-flip is \emph{not} injected---we
justify this in the next paragraph.  On each side the phase-flip
rate has four contributions,
\begin{equation*}
p_Z^{(j)} = p_{\rm loss} + p_{Z,\varphi}(\bar{n})
+ p_{\mathrm{pf,gate}}^{(j)}(\bar{n}) + p_{\mathrm{pf,Kerr}}^{(j)},
\end{equation*}
namely photon-loss dephasing $p_{\rm loss} = \bar{n}\,T_{\rm CX}/T_1$;
displaced-basis pure dephasing $p_{Z,\varphi}$ from $1/f$ noise
under the echo (Sec.~\ref{sec:sm-displaced-dephasing}); the coherent
gate-intrinsic residual $p_{\rm pf,gate}^{(j)}(\bar{n})$, log-space
interpolated from the same simulation used for the bit-flip; and
the echoed control self-Kerr contribution
$p_{\rm pf,Kerr}^{(j)}\sim(K_a\bar{n}\,T_{\rm CX})^4$
(Sec.~\ref{sec:dispecho_numerics}).
Idle errors on qubits not participating in the active operation of a
given schedule timestep are modeled as single-qubit $Z$ channels with
rate $\bar{n}\,T_{\rm slot}/T_1$, where $T_{\rm slot}$ is the slot
duration ($T_{\rm meas}$ for the measurement/reset slot,
$T_{\rm CX}$ for each of the two CNOT slots).

\emph{Why coherent ancilla bit-flips do not propagate.}
We do \emph{not} restabilize the ancilla between the two CNOT steps:
throughout the round the stabilization is off and the ancilla lives
in the displaced basis as $|0\rangle$ or $|2\alpha_c\rangle$.  The
data cat's conditional $X$ only depends on the coarse Fock-content
discrimination ``$n_c$ near zero'' vs ``$n_c \gg 0$'' seen by the
bin's cross-Kerr detuning $\chi\hat n_c$; whether the beam-splitter
drive is resonant or off-resonant is all that matters.  We therefore
sort ancilla errors during the round by how they act on
$\langle\hat n_c\rangle$:
\begin{enumerate}
\item[(i)] \emph{Coherent phase-space rotations of $|2\alpha_c\rangle$}
  (from control self-Kerr, echo imperfection, or residual cavity
  dephasing) preserve $\langle\hat n_c\rangle$ exactly, so the
  beam-splitter regime is unchanged during the round.  They only
  crystallize into a cat-basis bit-flip at end-of-round
  restabilization, when the rotated $|2\alpha_c\rangle$ is projected
  onto the wrong cat pole---after both CNOT steps are complete and
  no CX remains for the bit-flip to propagate through.
\item[(ii)] \emph{Single-photon incoherent events on $|2\alpha_c\rangle$}
  ($T_1$ loss, thermal heating) shift $\langle\hat n_c\rangle$ by
  $\mathcal{O}(1)$ against a $4|\alpha_c|^2$ separation, which does
  not switch the regime either; the resulting ancilla bit-flip is a
  direct photon-loss event on the ancilla and enters the error
  budget only through its direct action.
\item[(iii)] \emph{Branch-switching events} (cat bit-flips of the
  ancilla itself) are exponentially suppressed in $|\alpha_c|^2$
  because the two branches are separated by
  $\lvert\langle 0|2\alpha_c\rangle\rvert^2 = e^{-4|\alpha_c|^2}$.
\end{enumerate}
Classes~(i)--(iii) together justify keeping only the analytic
photon-loss contribution in $p_X^{(c)}$ above: (i) is confined to a
single post-round bit-flip channel that does not propagate; (ii) is
exactly the photon-loss term already retained; (iii) is
exponentially small.

\emph{Self-Kerr under echo timing.}
The displacement echo doubles the CNOT time
($T_{\rm CX} = 2\,T_{\rm cSWAP^2}$)---a $2\times$ increase in
photon-loss dephasing---in exchange for suppressing the control
self-Kerr phase-flip from $(K_a\bar{n}\,T)^2$ to
$(K_a\bar{n}\,T)^4$; for $\chi/K = 1000$ the resulting rate is
$\lesssim 10^{-7}$ per gate.  The interpolated per-gate rates above
come from a single-CNOT simulation (one cSWAP$^2$ plus its echo);
in the two-step syndrome round the ancilla accumulates self-Kerr
across \emph{both} steps before returning to vacuum, so deferring
the echo to end-of-round doubles the effective $T$ in the quartic
formula [Eq.~\eqref{eq:pZ_echo}] and inflates this contribution by
$\sim 16\times$.  Even
then it stays $\lesssim 10^{-5}$, negligible next to photon-loss
dephasing.

\subsection{Comparison with a Zeno CNOT in the same decoder}
\label{sec:sm-zeno-qec-comparison}

To compare against a Zeno-based CNOT under the same logical
decoder assumptions, we repeated the repetition-code simulation above
after replacing only the active-CNOT error inputs.  The code geometry,
syndrome-extraction schedule, idle-channel model, measurement/reset
error, storage-$T_1$ assumption, and MWPM decoder are identical to
those used in Fig.~\ref{main-fig:performance}(b).  For the Zeno CNOT
we use the optimized
phase-flip model from Refs.~\cite{Gautier2023zeno,LeRegent2023},
\begin{align}
p_{Z_a} &= \bar{n}\kappa_1 T
          + \frac{0.159}{\bar{n}\kappa_2 T}, \\
p_{Z_d} &= p_{Z_a Z_d}
          = \frac{1}{2}\bar{n}\kappa_1 T,
\end{align}
with $T$ chosen at the optimum
$T^\star = 0.282/(\bar{n}\sqrt{\kappa_1\kappa_2})$.  We take a
representative optimistic stabilization rate
$\kappa_2/(2\pi)=1$~MHz for the Zeno gate and use the same storage
$T_1$ values as in Fig.~\ref{main-fig:performance}(b).  This comparison is
therefore not a microscopic simulation of the Zeno gate dynamics; it is
an apples-to-apples insertion of the standard optimized Zeno CNOT error
rates into the same repetition-code decoder used for the VCB gate.

\begin{figure}[t]
\includegraphics[width=\columnwidth]{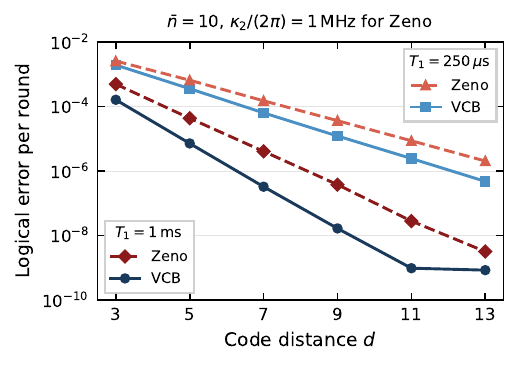}
\caption{\label{fig:vcb-vs-zeno}
Logical-error comparison between the VCB CNOT and an optimized
Zeno CNOT under the same repetition-code decoder and
syndrome-extraction schedule used in Fig.~\ref{main-fig:performance}(b).
The Zeno CNOT uses the phase-flip model from
Refs.~\cite{Gautier2023zeno,LeRegent2023} with
$\kappa_2/(2\pi)=1$~MHz, while the VCB curve uses the simulated
per-CNOT rates specified in Sec.~\ref{sec:sm-noise-model}.  The
storage-$T_1$, measurement/reset model, and idle channels are
matched between the two curves.
}
\end{figure}

Figure~\ref{fig:vcb-vs-zeno} shows that, at matched storage lifetime,
the VCB curve lies below the corresponding Zeno curve for this
representative Zeno operating point.  The difference comes primarily
from the optimized Zeno CNOT's non-adiabatic phase-flip contribution,
whereas the VCB noise budget in this regime is dominated by storage
loss accumulated during gates and measurement-cycle slots.  The precise
Zeno curve depends on
the assumed achievable $\kappa_2$ and on Zeno-specific leakage
mechanisms not modeled as separate correlated events here, so the plot
should be read as a same-simulation benchmark rather than a complete
architecture-level resource comparison.

\subsection{Why simultaneous two-target beam-splitting does not work}

A natural question is whether both data cats neighboring a given ancilla
could be coupled to the bin \emph{simultaneously}, reducing the gate
count from two sequential CNOTs to a single ``CXX'' operation.  However,
the simultaneous Hamiltonian
$\hat{H} = g(\hat{b}^\dagger\hat{a}_L + \hat{b}^\dagger\hat{a}_R + \mathrm{h.c.})$
couples only the symmetric (bright) normal mode
$(\hat{a}_L + \hat{a}_R)/\sqrt{2}$ to the bin; the antisymmetric (dark)
mode $(\hat{a}_L - \hat{a}_R)/\sqrt{2}$ is completely decoupled.
In the cat $Z$ basis, SWAP$^2$ on the bright mode swaps only the
same-sign states $|00\rangle_{LR}\leftrightarrow|11\rangle_{LR}$ while
leaving $|01\rangle_{LR}$ and $|10\rangle_{LR}$ unchanged.  This
implements a conditional swap within the even-parity subspace rather
than the $X_L\otimes X_R$ required for $ZZ$ stabilizer measurement.
Consequently, when the data pair is in a $Z$-eigenstate other than
$|00\rangle_{LR}/|11\rangle_{LR}$ (for example
$|01\rangle_{LR}$)---as it will be whenever the code has \emph{any}
$Z$ syndrome to detect---the vacuum-conditional operation leaves the
data untouched, so the ancilla learns nothing about that data pair
and the intended $ZZ$ stabilizer readout fails.

\section{Simulation methods and convergence}
\label{sec:sm-simulations}
All gate-level simulations in this work use the QuTiP~5 open-source
framework~\cite{qutip} to integrate the master equation (or, in the
absence of dissipation, the Schr\"{o}dinger equation) for the three-mode
system comprising the control cat, bin, and target cat modes.
The QEC simulations [Fig.~\ref{main-fig:performance}(b,c)] use the Stim circuit-level noise
simulator~\cite{stim} with Sinter~\cite{sinter} for statistical collection
and PyMatching~\cite{pymatching} for minimum-weight perfect matching
decoding.

\subsection{Hamiltonian and pulse parameters}

The time-dependent Hamiltonian is
\begin{equation}\label{eq:H_sim}
\hat{H}(t)
= \chi\,\hat{a}_c^\dagger\hat{a}_c\,\hat{b}^\dagger\hat{b}
  + \hat{H}_{\mathrm{Kerr}}
  + g(t)\bigl(\hat{b}^\dagger\hat{a}_t
    + \hat{b}\,\hat{a}_t^\dagger\bigr),
\end{equation}
where $\hat{H}_{\mathrm{Kerr}}$ collects quadratic self-Kerr terms
$(K_a/2)\,\hat{a}_c^{\dagger 2}\hat{a}_c^{2}$,
$(K_b/2)\,\hat{b}^{\dagger 2}\hat{b}^{2}$, and, where noted,
higher-order ($K_3$ through $K_6$) Kerr corrections.
The beam-splitter coupling envelope $g(t)$ is the Gaussian pulse
defined in Sec.~\ref{sec:sm-drag}.  In the simulations, the peak
amplitude $g_{\max}$ is calibrated numerically so that the pulse area
equals $\pi/2$ (each half of the $\chi$-echo pair) or $\pi$
(single-pulse mode).  For $\chi$-echo simulations, we use the
two-pulse sequence defined in Sec.~\ref{sec:sm-chiecho}.

\subsection{ODE solver}

We use the default adaptive-step ODE solver in
QuTiP~5 (\texttt{mesolve}) with absolute tolerance
$\epsilon_{\mathrm{abs}} = 10^{-13}$ and relative tolerance
$\epsilon_{\mathrm{rel}} = 10^{-11}$.  The time grid is sampled at
$10^3\,T^{-1}$ points per pulse half, providing dense output for
expectation-value monitoring.  The solver stores only the final state to
minimize memory usage; edge-state population on the highest Fock level of
each mode is monitored as a truncation diagnostic.

\subsection{Hilbert space dimensions}

The three-mode Hilbert space
$\mathcal{H} = \mathcal{H}_c \otimes \mathcal{H}_b \otimes
\mathcal{H}_t$ is truncated at mode-dependent Fock cutoffs
$[N_c, N_b, N_t]$.  The cutoffs are chosen adaptively: For each
$|\alpha|$, $N_t$ is set to the smallest integer such that the Fock-state
coefficient $|c_n(\alpha)|^2 = e^{-|\alpha|^2}|\alpha|^{2n}/n!$ falls
below a tolerance $\epsilon_{\mathrm{trunc}}$ at $n = N_t$.  Since the
control operates in a displaced frame at amplitude $2\alpha_c$ during
phase-flip simulations, $N_c$ is sized to accommodate
$|2\alpha_c\rangle$.  For the bit-flip ON branch (control in vacuum), we
set $N_c = 2$; for the OFF branch (control in $|2\alpha\rangle$), we use
the adaptive cutoff with $N_b = 4$.

The dimensions used for the two phase-flip panels
Fig.~\ref{main-fig:buildup}(d,e) are listed in Table~\ref{tab:dims},
at $\chi/2\pi = 4$~MHz and $\epsilon_{\mathrm{trunc}} = 10^{-2}$.
The three bit-flip panels (a--c) use the smaller cutoffs specified
above ($N_c = 2$ ON, adaptive $N_c$ with $N_b = 4$ OFF).

\begin{table}[h]
\caption{\label{tab:dims}
Hilbert space dimensions $[N_c, N_b, N_t]$ used in the phase-flip
$\alpha$-sweep of Fig.~\ref{main-fig:buildup}.
Total dimension is $N_c \times N_b \times N_t$.}
\vspace{16pt}
\begin{tabular}{c c c}
\hline\hline
$|\alpha|^2$ & $[N_c, N_b, N_t]$ & Total dim.\ \\
\hline
2  & $[15, 6, 6]$   & $540$   \\
4  & $[25, 9, 9]$   & $2{,}025$  \\
6  & $[34, 12, 12]$ & $4{,}896$  \\
8  & $[43, 15, 15]$ & $9{,}675$  \\
10 & $[52, 17, 17]$ & $15{,}028$ \\
\hline\hline
\end{tabular}
\end{table}

\subsection{Convergence}

We verified convergence of the error rates with
respect to three independent parameters:
(i)~\emph{Fock-space truncation:} we reduced
$\epsilon_{\mathrm{trunc}}$ until the extracted bit-flip errors
plateaued at all $|\alpha|^2$ values;
(ii)~\emph{ODE tolerances:} we tightened
$(\epsilon_{\mathrm{abs}}, \epsilon_{\mathrm{rel}})$ from
$(10^{-10}, 10^{-8})$ down to $(10^{-13}, 10^{-11})$, at which point
the extracted error probabilities plateau near the numerical floor
of ${\sim}10^{-9}$;
(iii)~\emph{time-grid density:} doubling the number of output points
produces no change at machine precision.
Additionally, the top-level Fock-state populations
$|\langle N_j - 1 | \rho | N_j - 1 \rangle|$ remain below $10^{-3}$
throughout the evolution for all modes, confirming that truncation
artifacts are negligible.

\subsection{QEC simulation implementation}

The circuit-level repetition-code noise model is defined in
Sec.~\ref{sec:sm-simcnot}.  For the logical error rate data in
Fig.~\ref{main-fig:performance}(b,c), each shot simulates $5d$ syndrome
rounds to ensure reliable error-rate extraction.  Statistical
collection uses up to $10^7$ shots per task with early stopping at
$10^3$ detected errors; logical error rates per round are computed via
the Bayesian estimator
$p_L = \mathrm{Beta}(1 + n_{\mathrm{err}},\,
1 + n_{\mathrm{shots}} - n_{\mathrm{err}})$ and converted to per-round
rates using the shot-to-piece decomposition.

The total logical error per round is the sum of the sinter-measured
$Z$-type error and an $X$-type contribution added analytically as
$d\,\kappa\,\bar{n}\,(2T_{\rm CX} + T_{\rm meas})\,e^{-2\bar{n}}$
(exponentially small in $\bar{n}$ and below the sampling sensitivity
of a $10^{7}$-shot run).
Fig.~\ref{main-fig:performance}(c) uses the same per-point pipeline
in an outer sweep over a $9\times15$ grid of
$\bar{n}\in\{2,3,\dots,10\}$ and $T_1$ log-spaced from $100$~\textmu s
to $10$~ms at fixed $d=11$.

\section{Toffoli gate extension}
\label{sec:sm-toffoli}
The vacuum-conditional beam-splitter mechanism extends naturally to a
bias-preserving Toffoli (CCX) gate by coupling \emph{two} control cat
qubits to a shared bin mode.  This provides a native non-Clifford gate
for the cat qubit architecture, completing a universal bias-preserving
gate set $\{Z(\theta), X, \text{CX}, \text{CCX}\}$ for
fault-tolerant quantum computation~\cite{Guillaud2019, Chamberland2022}.

\subsection{Two-control Hamiltonian}

Consider two control cats with annihilation operators $\hat{a}_{c1}$
and $\hat{a}_{c2}$, a shared bin mode $\hat{b}$, and a target cat
$\hat{a}_t$.  The two-control Hamiltonian is
\begin{equation}\label{eq:H_toffoli}
\hat{H}_{\rm CCX} =
  \chi_1\,\hat{a}_{c1}^\dagger\hat{a}_{c1}\,\hat{b}^\dagger\hat{b}
+ \chi_2\,\hat{a}_{c2}^\dagger\hat{a}_{c2}\,\hat{b}^\dagger\hat{b}
+ g(t)\bigl(\hat{b}^\dagger\hat{a}_t + \hat{b}\,\hat{a}_t^\dagger\bigr),
\end{equation}
where $\chi_1$ and $\chi_2$ are cross-Kerr couplings between each
control and the bin.  Each $\{\hat a_{cj}, \hat b\}$ pair shares the
same nonlinear properties as the single control--bin pair of the CNOT
gate, so the same error-suppression protocol applies to each control
independently.

\subsection{Vacuum-conditional truth table}

The key observation is that the beam-splitter drive at frequency
$\omega_b - \omega_t$ is resonant with the bin only when
\emph{both} control cats are in the vacuum state $|0\rangle$.  When
either (or both) controls are in $|2\alpha\rangle$, the cross-Kerr
shifts detune the bin:

\begin{center}
\renewcommand{\arraystretch}{1.2}
\begin{tabular}{c c c c}
\hline\hline
$c_1$ & $c_2$ & Avg.\ bin detuning $\bar{\Delta}$ & Action on target \\
\hline
$|0\rangle$ & $|0\rangle$ & $0$ & SWAP$^2$ ($X$ gate) \\
$|0\rangle$ & $|2\alpha\rangle$ & $4\chi_2|\alpha|^2$ & Identity \\
$|2\alpha\rangle$ & $|0\rangle$ & $4\chi_1|\alpha|^2$ & Identity \\
$|2\alpha\rangle$ & $|2\alpha\rangle$ & $4(\chi_1{+}\chi_2)|\alpha|^2$ & Identity \\
\hline\hline
\end{tabular}
\end{center}

\noindent In the cat logical basis
$\{|0_L\rangle = |\mathcal{C}_+\rangle,\;|1_L\rangle = |\mathcal{C}_-\rangle\}$
and the displaced control basis, the truth table above implements
$X_t$ conditioned on both controls being in $|0_L\rangle$.  With
appropriate single-qubit frame corrections (bit flips on the controls
before and after, or equivalently a basis relabeling), this yields the
standard Toffoli truth table:
\begin{equation}\label{eq:toffoli_truth}
\text{CCX}:\quad
|1\rangle_{c1}|1\rangle_{c2}|x\rangle_t
\;\to\; |1\rangle_{c1}|1\rangle_{c2}|x \oplus 1\rangle_t,
\end{equation}
with the target unchanged for all other control states.

\subsection{Bias preservation}

The bias-preservation argument follows the same logic as for the CNOT
gate:

\begin{enumerate}
\item \textbf{Target bit-flip:} The SWAP$^2$ preserves the
  coherent-state structure of the target, so the bit-flip error
  remains exponentially suppressed as $\sim\alpha\,e^{-2\alpha^2}$
  (Sec.~\ref{sec:sm-bitflip}).  The two controls do not affect this
  argument since they act only through the detuning of the bin.

\item \textbf{Control bit-flips:} In the ON branch (both controls in
  vacuum), $\hat{n}_{c1} = \hat{n}_{c2} = 0$, so both cross-Kerr
  terms vanish and the controls experience no photon-number-dependent
  forces.  In the OFF branches, the beam-splitter is far-detuned and
  the transient bin excitation $\bar{n}_b^{\rm peak} \sim (2g/\bar{\Delta}_{\rm min})^2$
  is suppressed by the \emph{smallest} average detuning
  $\bar{\Delta}_{\rm min} = 4\min(\chi_1,\chi_2)|\alpha|^2$.  This gives
  exponential suppression identical to the CNOT case, provided both
  $\chi_1$ and $\chi_2$ are sufficiently large.

\item \textbf{Phase-flips:} The same control phase-flip channels as
  for the CNOT (Secs.~\ref{sec:sm-dispecho} and~\ref{sec:sm-ctrl-pf})
  apply independently to each control, and the $\chi$-echo protocol
  can be extended to echo both $\chi_1$ and $\chi_2$ simultaneously
  by flipping both cross-Kerr couplings at the midpoint.
\end{enumerate}

\subsection{Error budget sketch}

The dominant errors for the Toffoli gate are:

\begin{itemize}
\item \textbf{Phase-flips (all modes):} We expect each control to
  inherit the same phase-flip budget as the CNOT control in
  Fig.~\ref{main-fig:buildup}, and the target to inherit that of the
  CNOT target.

\item \textbf{All bit-flip channels:} Exponentially suppressed as for
  the CNOT, $p_X \lesssim 10^{-7}$ at $|\alpha|^2 = 10$.
\end{itemize}

\noindent The total Toffoli error is therefore dominated by
the two control phase-flips,
$p_{\rm CCX} \approx 2p_Z^{(c)} + p_Z^{(t)}$,
comparable to the CNOT error at the same operating point.

\subsection{Connection to universal fault-tolerant quantum computation}

Refs.~\cite{Guillaud2019, Chamberland2022} show that the gate set
$\{Z(\theta),\, X,\, \text{CX},\, \text{CCX}\}$, together with
$|+\rangle$-state preparation and $X$-basis measurement, is universal
for quantum computation.  The subset $\{Z(\theta),\, X,\, \text{CX}\}$
alone is not, and the standard route to non-Clifford gates via a
bias-preserving Hadamard is not available: $H$ maps $Z \to X$ and
therefore cannot preserve the bias structure.  A native,
bias-preserving Toffoli sidesteps this by supplying the missing
non-Clifford element directly~\cite{Guillaud2019}.

Combining the CNOT of this work with the Toffoli extension, the full
bias-preserving gate set becomes:
\begin{itemize}
\item $Z(\theta)$: Zeno gate~\cite{Mirrahimi2014, Guillaud2019}
\item $X$: SWAP$^2$ beam-splitter (single-qubit, no ancilla needed)
\item CX: vacuum-conditional beam-splitter (this work)
\item CCX: two-control vacuum-conditional beam-splitter (this section)
\end{itemize}
All four gates are bias-preserving, enabling a repetition code that
corrects only the dominant (phase-flip) errors while the exponentially
suppressed bit-flip errors are handled by the bias.  A detailed
resource analysis of how the native CCX reduces magic state
distillation overhead relative to $T$-gate--based
synthesis~\cite{Selinger2013, Beverland2020} is left to future work.

\putbib[sm]
\end{bibunit}
\makeatother
\endgroup

\end{document}